\documentclass{article} 

\PassOptionsToPackage{table}{xcolor}
\newcommand{\revised}[1]{\textcolor{black}{#1}}

\usepackage{iclr2027_conference,times}

\usepackage{xcolor}
\definecolor{candorrow}{gray}{0.94}

\usepackage{amsmath,amsfonts,bm}

\def\eqref#1{equation~\ref{#1}}

\def\1{\bm{1}}

\DeclareMathAlphabet{\mathsfit}{\encodingdefault}{\sfdefault}{m}{sl}
\SetMathAlphabet{\mathsfit}{bold}{\encodingdefault}{\sfdefault}{bx}{n}

\usepackage{wrapfig}
\usepackage{needspace}
\usepackage{algorithm}
\usepackage{algorithmic}
\usepackage{amsmath,amssymb,bm}

\usepackage{hyperref}
\usepackage{url}
\usepackage{graphicx}
\usepackage{placeins}
\usepackage{booktabs}
\usepackage{adjustbox}
\usepackage[normalem]{ulem}
\usepackage{hyperref}
\usepackage{multirow}
\usepackage{array}

\title{Decompose Dynamics Before Learning Dependencies in Spatiotemporal Systems} 

\author{%
\makebox[\textwidth][c]{%
\begin{tabular}{c}
Ziqi Wang, Daojiang Hu, Cheng Bao, Zhiwei Ling, Wenzhuo Qian,\\
Jiahui Zhai, Hailiang Zhao\\[8pt]
\normalfont Zhejiang University
\end{tabular}%
}%
}

\iclrfinalcopy 
\begin{document}

\maketitle
\fancyhead{}
\begin{abstract}
Relations in networked spatiotemporal systems are often learned from observations that entangle dynamics governed by different mechanisms, obscuring what evolves locally and how it propagates across nodes.
We introduce Component-Aware Network Dynamics with Ordered Relations (CANDOR), which decomposes local dynamics before learning their dependencies. CANDOR represents each trajectory through a persistent
background, gradual accumulation and release, and sparse shocks. Conditioned on these components, a delay-aware physical branch models edge and sample-dependent propagation over directed topology, while a
topology-unconstrained functional branch discovers latent dependencies from background dynamics. Context-adaptive fusion combines functional, forward-propagation, and reverse-support forecasts, with training objectives encouraging specialized and semantically consistent representations. Experiments on two traffic benchmarks and three long-horizon water-quality datasets span two distinct spatiotemporal systems: human-driven urban traffic and naturally evolving river water quality. {\color{black}CANDOR achieves the lowest aggregate normalized error across five datasets, outperforming the overall runner-up by $7.25\%$.} These results establish decomposition before dependency learning as an
effective principle for spatiotemporal representation learning.
\end{abstract}

\section{Introduction}
Spatiotemporal systems consist of networked entities whose states evolve locally while influencing one another over time. They arise in both human mobility networks and environmental systems, where understanding their dynamics supports applications ranging from traffic management to environmental monitoring \citep{kumar2024spatio}. The key challenge is to understand how local dynamics evolve and how nodes influence each other, both through physical connections and beyond them.

\begin{figure}[t]
    \centering
    \includegraphics[width=1\linewidth]{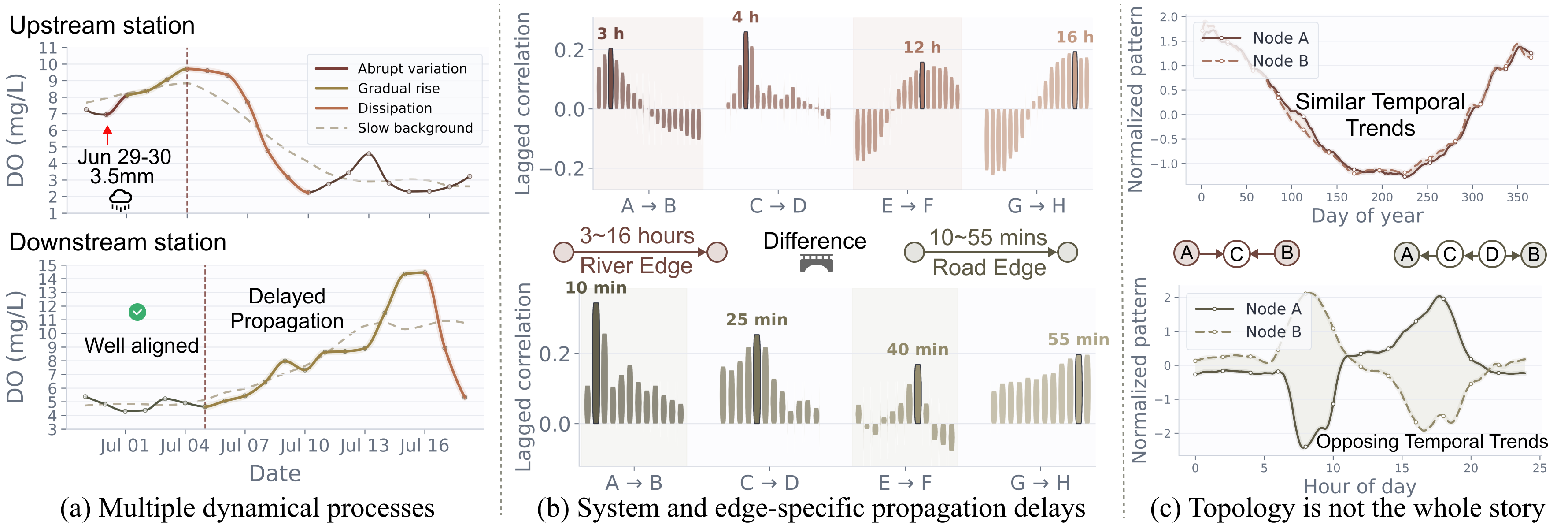}
    \caption{Representative characteristics of real-world spatiotemporal systems:
(a) coexisting dynamics, (b) edge and system-dependent delays reflected
by distinct peak lags, and (c) nonlocal dependencies beyond topology
with similar or opposing temporal patterns.}
\vspace{-1em}
    \label{fig1}
\end{figure}

Recent advances in spatiotemporal forecasting have enabled more flexible modeling of heterogeneous dynamics and adaptive spatial dependencies \citep{lee2024testam,lyu2025autostf,gao2025stssdl,zou2026metadg}. This progress reflects a broader shift from predefined spatial structures and shared temporal models toward learned relationships and context-dependent modeling \citep{qin2026adaptive}. While these developments increase the flexibility of dependency learning, they leave open a central question: how should different local dynamics inform the relationships learned between nodes? As illustrated in Figure~\ref{fig1}, local trajectories can combine persistent patterns, gradual changes, and abrupt shocks. Their effects may propagate through physical connections with delays that vary across systems and edges, while nodes without direct connections may also exhibit related temporal patterns. Learning dependencies from a single representation of these mixed dynamics can obscure how individual processes evolve, propagate, and relate across nodes. These observations motivate a central principle: \emph{decompose dynamics before learning dependencies}.

To this end, we introduce \textbf{Component-Aware Network Dynamics with Ordered Relations (CANDOR)}. CANDOR decomposes each trajectory into four components: background, accumulation, release, and shock, capturing persistent patterns, gradual buildup, gradual decline, and abrupt changes, respectively. Conditioned on this decomposition, a delay-aware physical branch captures interactions that vary across edges and samples over directed topology, while a functional branch learns latent dependencies beyond direct physical connections. CANDOR then adaptively fuses predictions from the functional, forward propagation, and reverse support branches. Its training objectives encourage branch specialization and semantic consistency between the decomposed components and their corresponding branch states at each node. Our contributions are threefold:

\begin{itemize}
    \item \textbf{Decomposition before dependency learning.} We introduce a component-aware formulation that disentangles local dynamics into background, accumulation, release, and shock, and uses the resulting components to guide the learning of dependencies between nodes.

    \item \textbf{Mechanism-specific relation learning.} We develop complementary physical and functional branches that capture delay-aware propagation over directed topology and latent dependencies beyond physical connectivity. Adaptive fusion integrates their predictions, while specialization objectives promote distinct and interpretable relational roles.

    \item \textbf{Cross-system validation.} We evaluate CANDOR on urban traffic and river water quality systems to assess its effectiveness across distinct spatiotemporal settings. We also release three water-quality datasets comprising over 1.8 million valid station-hour observations from 43 monitoring sites over five years, together with directed river-network information.
\end{itemize}

\section{Related Works}\label{sec:related}
{\color{black}
\paragraph{Dynamics Decomposition for Time Series Modeling.}
Recent decomposition-based methods separate temporal patterns to facilitate forecasting. TimeMixer \citep{wang2024timemixer} and TimeMixer++ \citep{wang2025timemixer++} mix seasonal and trend patterns across scales and resolutions, while TimeKAN models distinct frequency bands with specialized networks \citep{huang2025timekan}. Leddam combines learnable decomposition with inter-series and intra-series dependency modeling \citep{yu2024revitalizing}. Decomposition also guides component-specific masking in ST-MTM \citep{seo2025st} and separate seasonal and trend supervision in DBLoss \citep{qiu2026dbloss}. More recently, CEDAR separates action-conditioned state transitions from event-driven residual corrections \citep{meng2026cedar}, while a synthetic evaluation suite \citep{wu2026time} evaluates component recovery under controlled generative mechanisms and connects decomposition quality to downstream structure discovery. These developments extend decomposition from temporal pattern extraction to modeling distinct sources of variation and evaluating component fidelity. CANDOR uses decomposed dynamical roles to guide complementary mechanisms of cross-node relation learning, linking local evolution to network dependence.
}

{\color{black}
\paragraph{Spatiotemporal Dependency Modeling.}
General time-series forecasters capture temporal regularities through
convolutional, multiscale, and non-stationarity-aware architectures
\citep{moderntcn2024,pathformer2024,timebridge2025}. In networked
systems, a further challenge is identifying informative cross-node
dependencies under varying spatial and temporal contexts.
Graph-based methods address this challenge through structural priors,
adaptive connectivity, and dependency selection. PDFormer incorporates
propagation delays into spatial attention
\citep{jiang2023pdformer}. MegaCRN learns adaptive structures through
meta-graph learning \citep{jiang2023spatio}, while MetaDG jointly
generates dynamic adjacency and model parameters
\citep{zou2026metadg}. TESTAM routes temporal, static-graph, and
dynamic-graph experts according to context \citep{lee2024testam}.
TimeFilter selects patch-specific dependencies
\citep{hu2025timefilter}, while DUET learns cross-series associations
without predefined physical connectivity \citep{qiu2025duet}.
Complementing graph adaptation, signal decomposition provides a
perspective on distinguishing local evolution from cross-node
interactions. D$^2$STGNN develops this perspective by separating
diffusion and inherent dynamics. It models diffusion through
bidirectional spatiotemporally localized convolution with physical
and adaptive graph supports, while capturing inherent dynamics
through recurrent and attention-based temporal modeling
\citep{ShaoZWWXCJ22}.

CANDOR grounds complementary relational mechanisms in
reference-anchored dynamical roles. Background dynamics guide
topology-unconstrained functional relations, while shock-modulated
states propagate over directed links with temporal-offset-specific
and sample-conditioned edge affinities. Role-consistency objectives
connect accumulation and release to propagation states, while adaptive fusion balances the contributions of complementary relational mechanisms.
}


\section{Methodology}
\subsection{Problem Formulation and Overview}

Let $\mathcal{G}_0=(\mathcal{V},\mathcal{E})$ denote a directed physical network with $N=|\mathcal{V}|$ nodes, where $(i,j)\in\mathcal{E}$ represents a connection from node $i$ to node $j$. Given historical observations $\mathbf{X}_{t-L+1:t}\in\mathbb{R}^{L\times N}$ and $\mathcal{G}_0$, and the ground-truth future $\mathbf{Y}^{\star}_{t+1:t+H}\in\mathbb{R}^{H\times N}$, the task is to predict $\mathbf{Y}_{t+1:t+H}\in\mathbb{R}^{H\times N}$, where $t$ is the current time, $L$ is the history length, and $H$ is the forecasting horizon.

\begin{figure*}[htbp]
\centering
\includegraphics[width=1\textwidth]{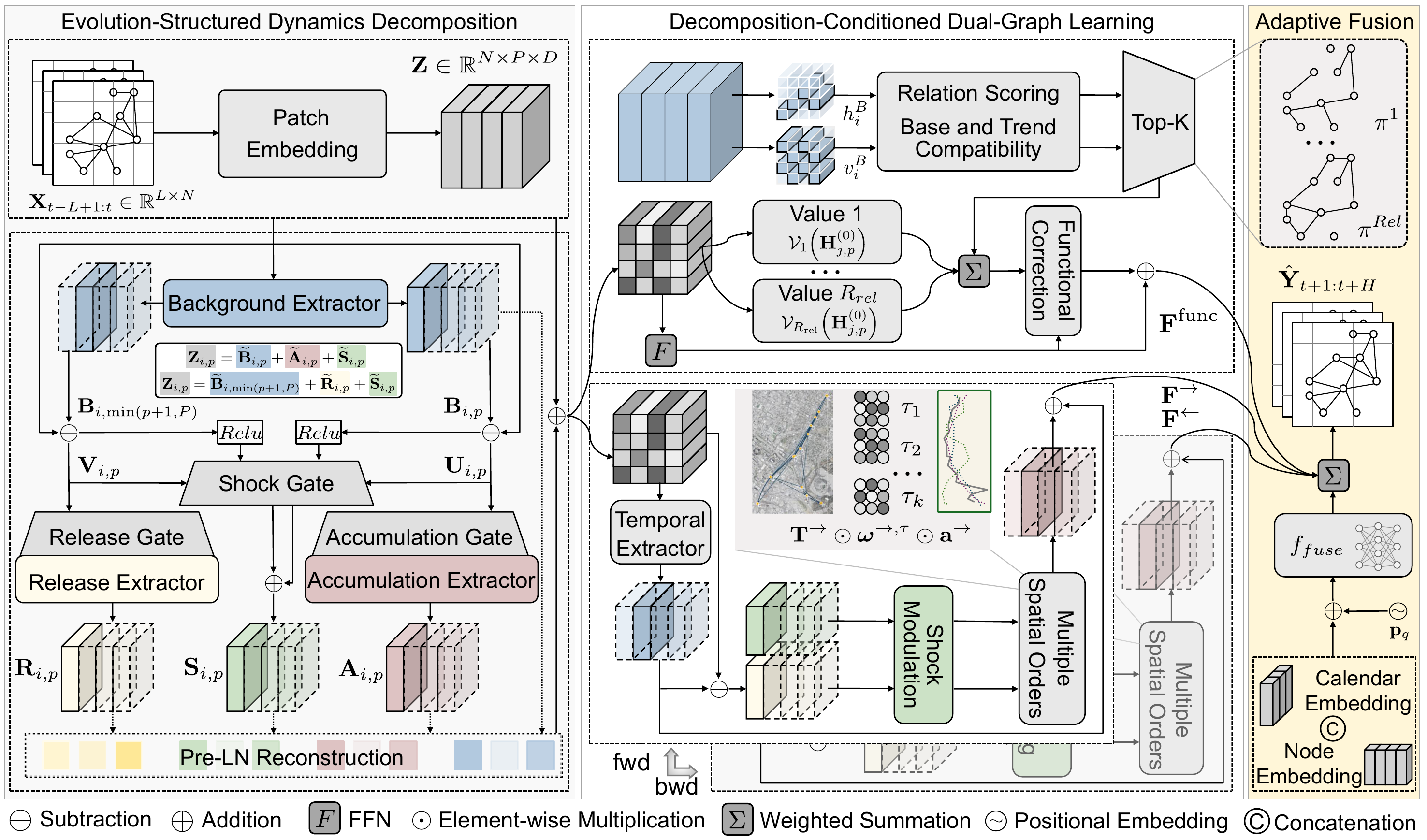}
\caption{\textbf{Overview of CANDOR.}
\textbf{(1) Evolution-structured dynamics decomposition} separates persistent and transient dynamics through complementary residual references.
\textbf{(2) Decomposition-conditioned dual-graph learning} combines delay-aware propagation over directed physical support with background-conditioned functional relations beyond topology.
\textbf{(3) Context-adaptive fusion} integrates functional, forward, and reverse-support forecasts contextually.}
\label{main}
\vspace{-0.6em}
\end{figure*}

CANDOR decomposes local dynamics before learning dependencies. As shown in Figure \ref{main}, it first separates node representations into four components: background, accumulation, release, and shock. These components inform two complementary relational mechanisms: the physical propagation graph captures delayed interactions over directed physical connections, while the functional relation graph uses background representations to infer dependencies beyond physical adjacency. CANDOR then adaptively fuses functional, forward-propagation, and reverse-support representations to generate the forecast. This organization connects local dynamics to complementary and mechanism-specific relations, avoiding dependency learning from entangled observations.


\subsection{Evolution-Structured Dynamics Decomposition}
\label{sec:decomposition}

\begin{figure*}[htbp]
\centering
\includegraphics[width=1\textwidth]{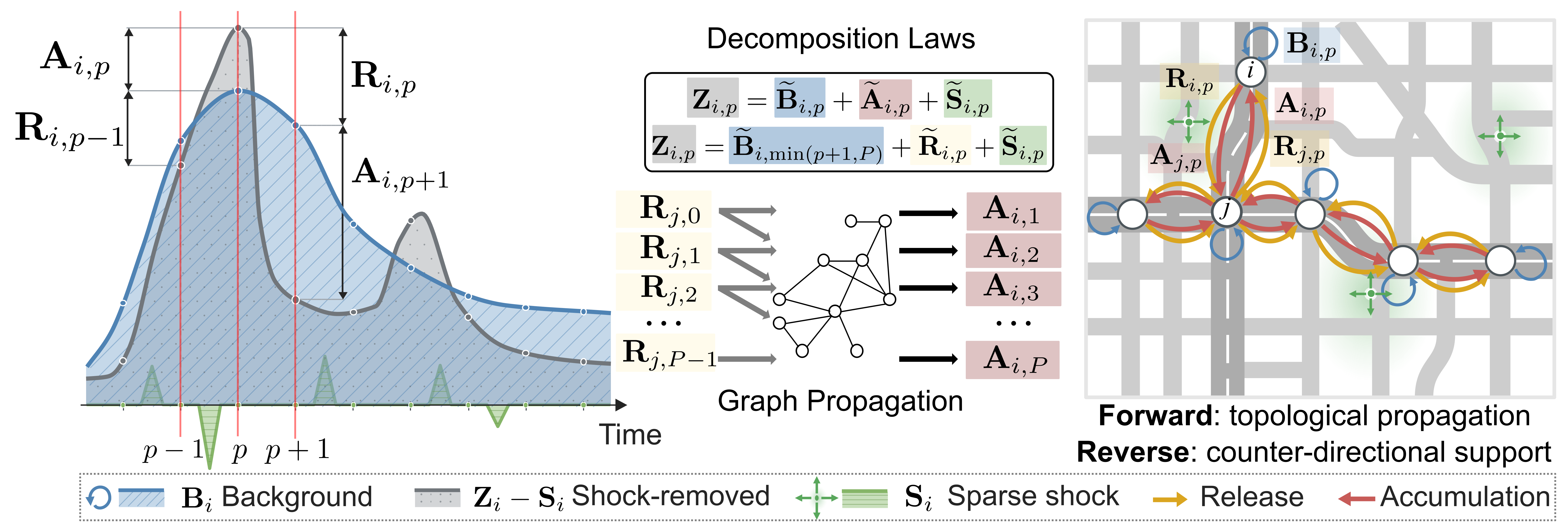}
\caption{\textbf{Evolution-structured dynamics decomposition.}
A shared background defines complementary residual views for accumulation and release, while sparse shocks capture abrupt deviations.
The components provide distinct local-evolution roles for relation learning.
Release-associated residuals propagate over the physical graph and support subsequent downstream accumulation.}
\label{decom}
\vspace{-0.6em}
\end{figure*}

{\color{black}
Spatiotemporal propagation systems couple local state evolution
with network transport and external forcing. For example, traffic
congestion forms and dissipates through imbalances between demand
and discharge \citep{daganzo1995cell}, while water-quality dynamics in river networks reflect production, consumption, exchange, and transport
\citep{sullivan2011modeling}. These mechanisms motivate distinguishing
persistent operating conditions, gradual buildup and relaxation, and
abrupt disturbances. Therefore, we organize local dynamics by both
their reference state and temporal evolution, accounting for changes
in the local baseline. The mechanistic basis is detailed in Appendix~\ref{app:decomposition_physics}.

We first embed the observed history as
$\mathbf{Z}=\operatorname{PatchEmbed}(\mathbf{X}_{t-L+1:t})
\in\mathbb{R}^{N\times P\times D}$,
where $P$ is the number of historical patches and $D$ is the latent
dimension. A causal temporal convolution $\mathcal{C}_{B}$ followed
by layer normalization $\operatorname{LN}_{B}$ extracts the background,
$\mathbf{B}=\operatorname{LN}_{B}(\mathcal{C}_{B}(\mathbf{Z}))$.
To account for this evolving reference, we construct two residual
views: the current-background view measures the present departure,
while the next-background view expresses the same state relative
to the subsequent baseline:
\begin{equation}
\mathbf{U}_{i,p}=\mathbf{Z}_{i,p}-\mathbf{B}_{i,p},
\qquad
\mathbf{V}_{i,p}
=\mathbf{Z}_{i,p}-\mathbf{B}_{i,\min(p+1,P)},
\label{eq:evolution_references}
\end{equation}
where $i$ indexes nodes and $p$ indexes historical patches.
Both references lie within the observed window, with the final
background repeated at the boundary. For $p<P$, their contrast
satisfies
$\mathbf{U}_{i,p}-\mathbf{V}_{i,p}
=\mathbf{B}_{i,p+1}-\mathbf{B}_{i,p}$,
explicitly capturing the shift in the local reference. We extract
accumulation $\mathbf{A}$ and release $\mathbf{R}$ from these views,
respectively, anchoring their roles to the current and subsequent
backgrounds. Figure~\ref{decom} illustrates the two views and their
connection to propagation. The shared background defines accumulation and release relative to the current and subsequent reference states, while sparse shocks account for abrupt deviations in both views.

First and second-order temporal differences characterize the
direction and variation of local changes, helping distinguish
buildup from recovery at similar state levels. Together with node
and calendar context, they guide adaptive component selection within
each view. A shared shock component $\mathbf{S}$ is extracted from
both views using residual and temporal-difference magnitudes to
select pronounced departures and rapid changes of either sign.
The resulting components provide structured inputs for relation
learning, with reconstruction and role-consistency objectives
reinforcing their intended roles during training
(Section~\ref{sec:learning_objectives}).
Full decomposition computations are provided in Appendix~\ref{app:decomposition}.

}

\subsection{Decomposition-Conditioned Dual-Graph Learning}
\label{sec:dual_graph}
We integrate the four components through a residual projection to obtain the refined history $\mathbf{H}^{(0)}\in\mathbb{R}^{N\times P\times D}$. The decomposition conditions the two branches in distinct ways: the physical branch propagates states derived from
$\mathbf{H}^{(0)}$ over directed topology, whereas the functional
branch additionally uses $\mathbf{B}$ to determine relational
weights. This gives the components distinct downstream roles:
the background additionally serves as a latent signature for
inferring non-topological functional relations, while accumulation,
release, and shock characterize transient dynamics involved in
propagation and state evolution.

{\color{black}
\paragraph{Physical propagation graph.}

Spatiotemporally localized propagation has been used to aggregate
bidirectional neighbors over a bounded temporal window
\citep{ShaoZWWXCJ22}. Building on this formulation, our physical
branch combines shock-conditioned propagation states with
temporal-position-specific and sample-conditioned affinities
along directed connections.
Starting from $\mathbf{H}^{(0)}$, we maintain separate
forward-propagation and reverse-support paths
$d\in\{\rightarrow,\leftarrow\}$.
Each path first extracts a node-wise temporal state
$\mathbf{L}^{d}$ from the history gated by node and calendar context
and removes its temporal backcast to obtain a
propagation residual $\mathbf{E}^{d}$.
To account for abrupt local disturbances, we further
introduce shock-conditioned modulation to adaptively
refine the propagation features.
The shock context $\mathbf{S}^{\mathrm{ctx}}$ is initialized
from the pre-normalization shock representation
$\widetilde{\mathbf{S}}$ for observed patches and
recursively estimated from previously available
contexts during future forecasting.
The resulting propagation states are formulated as
\begin{equation}
\mathbf{E}^{d}=\operatorname{LN}_{d}\!\left(\widetilde{\mathbf{H}}^{d}-\operatorname{ReLU}(\mathcal{P}_{\mathrm{back}}^{d}(\mathbf{L}^{d}))\right),
\quad
\mathbf{E}_{\mathrm{used}}^{d}=\mathbf{E}^{d}+\Delta_{S}^{d}(\mathbf{E}^{d},\mathbf{S}^{\mathrm{ctx}}),
\label{eq:physical_states}
\end{equation}
where $\widetilde{\mathbf{H}}^{d}$ denotes the history representation gated by node and calendar context,
$\mathcal{P}_{\mathrm{back}}^{d}$ is the learned temporal
backcast projection, and $\Delta_{S}^{d}$ represents
direction-specific shock modulation.
The modulated state $\mathbf{E}_{\mathrm{used}}^{d}$
is used for spatial aggregation, while the original
residual $\mathbf{E}^{d}$ is retained for
role-consistency supervision.

To capture heterogeneous interactions across temporal
positions and observed samples, we combine directed
physical support with position-specific and sample-conditioned affinities.
For a localized temporal window of $K_t$ positions
ending at patch $p$, the direction-specific
propagation operator and first-order graph
contribution are defined as
\begin{equation}
\mathbf{W}^{d,\tau}=\operatorname{RowNorm}(\mathbf{T}^{d}\odot\boldsymbol{\omega}^{d,\tau}\odot\mathbf{a}^{d}),
\quad
\mathbf{g}_{i,p}^{d,[1]}=\sum_{\tau=1}^{K_t}\sum_{j\neq i}W_{ij}^{d,\tau}\boldsymbol{\Psi}_{j,p,\tau}^{d},
\label{eq:physical_graph}
\end{equation}
where $\boldsymbol{\Psi}_{j,p,\tau}^{d}$ denotes
the transformed temporal-window feature derived
from $\mathbf{E}_{\mathrm{used}}^{d}$.
The directional support $\mathbf{T}^{d}$ is
constructed from the directed physical graph
$\mathcal{G}_0$.
The affinity $\boldsymbol{\omega}^{d,\tau}$
captures edge and temporal-position preferences shared across samples, whereas
$\mathbf{a}^{d}$ adapts to the observed history
and calendar context of each sample and remains
shared across temporal positions.
Sample indices are omitted throughout for clarity.
$\odot$ denotes element-wise multiplication, and
$\operatorname{RowNorm}$ normalizes over source nodes
while preserving zero rows.
To capture interactions along longer propagation
paths, we further aggregate spatial orders
$m=1,\ldots,K_s$, where $K_s$ is the maximum order.
The local temporal summary and multi-order graph
contributions are concatenated and projected to obtain
the directional propagation contribution $\mathbf{o}_{i,p}^{d}$.

Each path predicts the next latent patch state
by combining its temporal state with the incoming
graph contribution.
The two predictions are then coupled
through a shared latent feedback state defined by
\begin{equation}
\widehat{\mathbf{h}}_{i,p+1}^{d}=\mathbf{L}_{i,p}^{d}+\mathbf{o}_{i,p}^{d},
\quad
\overline{\mathbf{h}}_{i,p+1}=\tfrac12\left(\widehat{\mathbf{h}}_{i,p+1}^{\rightarrow}+\widehat{\mathbf{h}}_{i,p+1}^{\leftarrow}\right).
\label{eq:physical_rollout}
\end{equation}
The shared latent state is fed back to both paths
at every future patch, while their temporal
memories, propagation residuals, and graph
operators remain direction-specific.
The two directional prediction sequences are
retained separately and projected to
$\mathbf{F}^{\rightarrow}$ and
$\mathbf{F}^{\leftarrow}\in
\mathbb{R}^{N\times P_{\mathrm{out}}\times D_{\mathrm{f}}}$
for subsequent context-adaptive fusion, where
$P_{\mathrm{out}}$ is the number of target patches
and $D_{\mathrm{f}}$ is the forecast representation dimension.
Detailed state updates, shock modulation,
multi-order propagation, and coupled forecasting
are provided in Appendix~\ref{app:dual_graph}.
}

\paragraph{Functional relation graph.}
Functional dependencies can extend beyond physical topology
and arise from shared background dynamics (e.g., similar persistent temporal patterns among non-adjacent nodes). In this case, we construct the functional relation graph $\mathcal{G}_{\mathrm{func}}$ from background dynamics,
evaluating compatibility in both background state and evolution.
For node $i$, a descriptor $\mathbf{h}_{i}^{B}$ summarizes
background level and trajectory shape with node
and calendar context; a separate descriptor
$\mathbf{v}_{i}^{B}$ summarizes background trends.
Multiple relation channels capture different compatibility
patterns in these descriptors. For each channel $r$, we compute
\begin{equation}
s_{ij}^{r}
=
(\mathbf{h}_{i}^{B})^{\top}\mathbf{M}_{r}\mathbf{h}_{j}^{B}
+
\eta(\mathbf{v}_{i}^{B})^{\top}\mathbf{N}_{r}\mathbf{v}_{j}^{B}
+b_r,
\qquad
\pi_{ij}^{r}
=
\operatorname{TopKSoftmax}_{(j,r)}(s_{ij}^{r}),
\label{eq:functional_graph}
\end{equation}
where $\mathbf{M}_{r}$ and $\mathbf{N}_{r}$ are learned
relation-specific matrices, $b_r$ is a bias, and $\eta$
weights trend compatibility.
For each receiving node, $\operatorname{TopKSoftmax}$
jointly selects the top $k$ neighbor-relation pairs and
normalizes their scores without a physical adjacency mask.
This jointly determines which nodes contribute and through
which relation channels.

The weights $\pi_{ij}^{r}$ aggregate relation-specific
transformations of $\mathbf{H}^{(0)}$ to refine the temporal
forecast, yielding
$\mathbf{F}^{\mathrm{func}}\in
\mathbb{R}^{N\times P_{\mathrm{out}}\times D_{\mathrm{f}}}$.
Background dynamics guide relation selection, while all four
components contribute to information exchange.
The three forecast representations ($\mathbf{F}^{\mathrm{func}}$, $\mathbf{F}^{\rightarrow}$,
and $\mathbf{F}^{\leftarrow}$) are then passed to
adaptive fusion. Full computations are provided in
Appendix~\ref{app:dual_graph}.

\subsection{Context-Adaptive Forecast Fusion}
\label{sec:fusion}
The relative contributions of functional dependence, forward
propagation, and reverse-support context can vary across
nodes and forecasting positions. Therefore, we condition
their fusion on node identity, target calendar information,
and relative forecast position, allowing the three mechanisms
to contribute differently across prediction contexts.

Let $q$ index target patches and
$\mathcal{B}=\{\mathrm{func},\rightarrow,\leftarrow\}$
denote the three branches.
We construct a context vector $\mathbf{c}_{i,q}$ by
concatenating a learned node embedding, target time-of-day
and day-of-week embeddings, and a sinusoidal encoding of $q$.
An MLP $f_{\mathrm{fuse}}$ maps this context to three logits,
which determine the fused representation:
\begin{equation}
\boldsymbol{\lambda}_{i,q}
=
\operatorname{softmax}
\!\left(f_{\mathrm{fuse}}(\mathbf{c}_{i,q})\right),
\qquad
\mathbf{F}^{\mathrm{fuse}}_{i,q}
=
\sum_{b\in\mathcal{B}}
\lambda_{i,q}^{b}\mathbf{F}^{b}_{i,q},
\label{eq:context_fusion}
\end{equation}
where $\lambda_{i,q}^{b}$ is the weight of branch $b$ and it is shared across its latent features.
Calendar context captures recurring temporal variation,
while positional encoding distinguishes forecasting positions
even when their calendar contexts coincide. A shared prediction head decodes $\mathbf{F}^{\mathrm{fuse}}$
into patch-level predictions, which are averaged over overlaps
to obtain the final forecast $\hat{\mathbf{Y}}_{t+1:t+H}$.
Further details on adaptive branch fusion and patch-level forecast decoding are provided in Appendix~\ref{app:fusion}.

\subsection{Learning Objectives}
\label{sec:learning_objectives}

Prediction supervision alone does not ensure that decomposed
components retain their intended roles during relation learning.
We constrain both how they explain local dynamics and how they align with physical propagation states.

Let $\widetilde{\mathbf{B}}$, $\widetilde{\mathbf{A}}$,
$\widetilde{\mathbf{R}}$, and $\widetilde{\mathbf{S}}$
denote component outputs before normalization.
We impose reconstruction constraints consistent with the
two background references:
\begin{equation}
\begin{aligned}
\mathcal{L}_{\mathrm{rec}}
={}&\frac{1}{2}\operatorname{SL1}\!\left(
\widetilde{\mathbf{B}}+\widetilde{\mathbf{A}}
+\widetilde{\mathbf{S}},\operatorname{sg}(\mathbf{Z})\right)\\
&+\frac{1}{2}\operatorname{SL1}\!\left(
\widetilde{\mathbf{B}}_{:,2:P}
+\widetilde{\mathbf{R}}_{:,1:P-1}
+\widetilde{\mathbf{S}}_{:,1:P-1},
\operatorname{sg}(\mathbf{Z}_{:,1:P-1})\right),
\end{aligned}
\label{eq:evolution_reconstruction}
\end{equation}
where $\operatorname{SL1}$ is mean smooth-$L_1$ loss and
$\operatorname{sg}$ stops gradients through its argument.
Accumulation explains a local state relative to its current
background, whereas release explains it relative to the next
background, consistent with the two residual references in Eq.~\ref{eq:evolution_references}. Shock contributes to both reconstructions.
These soft constraints give the two residual views distinct
learning targets.

We further introduce a role-consistency loss
$\mathcal{L}_{\mathrm{role}}$ to connect the decomposition
with physical-path states. At each node, projected cosine
similarity aligns background at the next patch with the
current temporal states, and accumulation at the next patch
with incoming graph contributions $\mathbf{o}^{\Sigma}$. Release is aligned with
the corresponding propagation residuals $\mathbf{E}^{\Sigma}$ at the same patch.
Forward and reverse states are combined for these comparisons.
This encourages the components to retain their intended
roles in local prediction and relational propagation. The overall training objective is
\begin{equation}
\mathcal{L}
=
\mathcal{L}_{\mathrm{pred}}
+\mathcal{L}_{\mathrm{aux}}
+\chi(e)\left(
\lambda_{\mathrm{rec}}\mathcal{L}_{\mathrm{rec}}
+\mathcal{L}_{\mathrm{role}}\right),
\label{eq:overall_objective}
\end{equation}
where $\mathcal{L}_{\mathrm{pred}}$ supervises the final forecast.
The weighted auxiliary objective $\mathcal{L}_{\mathrm{aux}}$
supervises each branch before fusion, enforces historical
prediction consistency, and discourages redundancy between
temporal and propagation states. It also encourages smooth
background evolution and selective shock activation.
The coefficient $\lambda_{\mathrm{rec}}$ weights reconstruction,
and $\chi(e)$ activates reconstruction and role consistency
at the dataset-specific epoch $e_0$.
Full definitions are provided in
Appendix~\ref{app:training_objectives}.

\section{Experiments}

\paragraph{Tasks and datasets.}
We evaluate CANDOR on two contrasting spatiotemporal systems:
human-driven urban traffic and natural river networks governed by
directed hydrologic transport. Traffic combines recurrent mobility
patterns with abrupt congestion, whereas river systems exhibit sparse
directional topology and delayed downstream effects, providing distinct
tests of local dynamics and cross-node propagation. For traffic forecasting, we use METR-LA \citep{jagadish2014big} and PEMS-BAY \citep{chen2001freeway}, containing 207
and 325 sensors, respectively, and predict 12 five-minute steps from 12
historical steps. We further introduce three five-year datasets of
hourly dissolved-oxygen observations from BASIN030501 (Santee),
BASIN031300 (Apalachicola), and BASIN180102 (Klamath). They
contain 1.88 million station-hour measurements from 43 monitoring sites
and 87 directed river connections. We use 48 historical hours to
forecast the following 240 hours. Chronological train/validation/test
splits are 70/10/20 for traffic and 60/20/20 for water quality.
Dataset construction and quality-control procedures are provided in
Appendix~\ref{app:data}.

\paragraph{Implementation details.}
CANDOR uses a hidden dimension of 32, a forecast representation
dimension of 256, and a dropout rate of 0.1, resulting in fewer than
$0.91$M parameters on every dataset. The physical branch uses a
two-patch propagation window and two spatial orders to capture delayed
and multi-hop propagation with limited overhead. The functional branch
retains the top 16 node-relation pairs for each receiving node from up
to four relation channels, encouraging sparse dependency learning.
Patch length and stride are $4/2$ for METR-LA and $3/3$ for PEMS-BAY;
for water quality, they scale with the forecast horizon and reach
$16/8$ from 192 hours onward. We train CANDOR with Adam at a learning
rate of $2\times10^{-3}$.
Complete hyperparameter settings, sensitivity analyses, and
time-complexity analysis are provided in
Appendix~\ref{app:implementation}.

\paragraph{Baselines and evaluation protocol.}

{\color{black}We compare CANDOR with 16 representative baselines from two
complementary model families.} This choice reflects the heterogeneity of spatiotemporal systems: graph-based models are advantageous when cross-node propagation is informative, whereas general time-series models can be more effective when spatial coupling is weak or noisy and node-wise temporal regularities dominate. {\color{black}Spatiotemporal forecasting methods include
TESTAM~\citep{lee2024testam},
AutoSTF~\citep{lyu2025autostf},
ST-SSDL~\citep{gao2025stssdl},
D$^2$STGNN~\citep{ShaoZWWXCJ22},
HyperD~\citep{shao2026hyperd},
RSTIB-MLP~\citep{rstib2025},
STPGNN~\citep{kong2024stpgnn},
MSTHG~\citep{msthg2026}, and
FaST~\citep{fast2026}.
They cover adaptive graph learning,
propagation-aware modeling, hypergraph dependencies, and multiscale spatiotemporal representations. General time-series forecasters include
DTAF~\citep{dtaf2026},
LatentTSF~\citep{latenttsf2026},
ModernTCN~\citep{moderntcn2024},
Pathformer~\citep{pathformer2024},
PhaseFormer~\citep{phaseformer2026},
TimeBridge~\citep{timebridge2025}, and
TimeMixer++~\citep{wang2025timemixer++}.}
These methods span convolutional, decomposition-based, frequency-aware, and multiscale architectures.
Including both families enables evaluation across systems with different degrees of propagation strength without assuming that spatial modeling is uniformly beneficial. Detailed descriptions for all baselines are provided in Appendix~\ref{methods}.

All methods use identical chronological splits, history lengths, forecast horizons, targets, and evaluation pipelines. Each experiment is repeated using the same three seeds. Traffic forecasting uses MAE as the primary metric, with RMSE and MAPE as complementary measures; water-quality forecasting uses MSE as the primary metric and MAE as a complementary measure. All metrics are computed on the original data scale. For readability, the main tables report the top eight baselines: traffic models are selected separately by unrounded overall MAE, while a shared water-quality set is ranked by MSE averaged across all basins and horizons. {\color{black}Complete results for all 16 baselines and all water-quality horizons, including mean$\pm$std over three seeds, are provided in Appendix~\ref{app:res}.}



\subsection{Main Results}

\begin{table}[htbp]
{\color{black}
\centering
\vspace{-4pt}
\caption{
Mean forecasting performance over three seeds on METR-LA and PEMS-BAY.
For each dataset, the eight strongest baselines are independently selected
from 16 candidates according to mean MAE. Best results are in \textbf{bold} and
second-best results are \underline{underlined}.
}

\label{tab:traffic_comparison}

\vspace{4pt}

\scriptsize
\setlength{\tabcolsep}{2.25pt}
\renewcommand{\arraystretch}{1}

\resizebox{\textwidth}{!}{
\begin{tabular}{@{}l|ccc|ccc|ccc|ccc@{}}
\toprule

\multicolumn{13}{c}{\textbf{METR-LA}} \\

\midrule

\multicolumn{1}{c|}{
    \raisebox{-0.55\baselineskip}{\textbf{Method}}
}
& \multicolumn{3}{c|}{\textbf{Overall}}
& \multicolumn{3}{c|}{\textbf{15 min}}
& \multicolumn{3}{c|}{\textbf{30 min}}
& \multicolumn{3}{c}{\textbf{60 min}} \\

& \textbf{MAE} & \textbf{RMSE} & \textbf{MAPE}
& \textbf{MAE} & \textbf{RMSE} & \textbf{MAPE}
& \textbf{MAE} & \textbf{RMSE} & \textbf{MAPE}
& \textbf{MAE} & \textbf{RMSE} & \textbf{MAPE} \\

\midrule

\revised{D$^2$STGNN~\citep{ShaoZWWXCJ22}}
& \revised{\underline{2.870}} & \revised{\underline{5.879}} & \revised{\underline{7.80\%}}
& \revised{\underline{2.554}} & \revised{\underline{4.898}} & \revised{\underline{6.52\%}}
& \revised{\underline{2.902}} & \revised{\underline{5.916}} & \revised{\underline{7.92\%}}
& \revised{\underline{3.343}} & \revised{\underline{7.033}} & \revised{\underline{9.64\%}} \\

TESTAM~\citep{lee2024testam}
& \revised{2.982} & \revised{6.136} & 8.14\%
& 2.670 & 5.206 & 6.89\%
& \revised{3.017} & \revised{6.189} & 8.25\%
& \revised{3.420} & \revised{7.215} & 9.86\% \\

ST-SSDL~\citep{gao2025stssdl}
& 2.986 & 6.191 & \revised{8.13\%}
& \revised{2.643} & \revised{5.133} & \revised{6.80\%}
& 3.021 & 6.225 & \revised{8.23\%}
& 3.474 & 7.429 & 10.00\% \\

AutoSTF~\citep{lyu2025autostf}
& 2.999 & 6.195 & 8.16\%
& 2.692 & 5.231 & 6.97\%
& 3.040 & 6.254 & 8.29\%
& 3.427 & 7.292 & \revised{9.81\%} \\

MSTHG~\citep{msthg2026}
& 3.129 & 6.357 & 8.84\%
& 2.790 & 5.388 & 7.42\%
& 3.177 & 6.445 & 9.03\%
& 3.602 & 7.460 & 10.71\% \\

HyperD~\citep{shao2026hyperd}
& 3.370 & 6.721 & 10.29\%
& 3.091 & 5.956 & 9.20\%
& 3.451 & 6.832 & 10.57\%
& 3.778 & 7.747 & 12.04\% \\

STPGNN~\citep{kong2024stpgnn}
& 3.485 & 6.888 & 10.12\%
& 2.977 & 5.706 & 8.12\%
& 3.499 & 6.911 & 10.18\%
& 4.257 & 8.339 & 13.08\% \\

RSTIB-MLP~\citep{rstib2025}
& 3.778 & 9.082 & 10.12\%
& 3.221 & 7.434 & 8.36\%
& 3.856 & 9.265 & 10.39\%
& 4.522 & 10.836 & 12.41\% \\

\midrule
\rowcolor{candorrow}
CANDOR (Ours)
& \textbf{2.854} & \textbf{5.860} & \textbf{7.77\%}
& \textbf{2.548} & \textbf{4.892} & \textbf{6.47\%}
& \textbf{2.889} & \textbf{5.891} & \textbf{7.86\%}
& \textbf{3.300} & \textbf{6.991} & \textbf{9.62\%} \\

\addlinespace[1pt]
\midrule

\multicolumn{13}{c}{\textbf{PEMS-BAY}} \\

\midrule

\multicolumn{1}{c|}{
    \raisebox{-0.55\baselineskip}{\textbf{Method}}
}
& \multicolumn{3}{c|}{\textbf{Overall}}
& \multicolumn{3}{c|}{\textbf{15 min}}
& \multicolumn{3}{c|}{\textbf{30 min}}
& \multicolumn{3}{c}{\textbf{60 min}} \\

& \textbf{MAE} & \textbf{RMSE} & \textbf{MAPE}
& \textbf{MAE} & \textbf{RMSE} & \textbf{MAPE}
& \textbf{MAE} & \textbf{RMSE} & \textbf{MAPE}
& \textbf{MAE} & \textbf{RMSE} & \textbf{MAPE} \\

\midrule

\revised{D$^2$STGNN~\citep{ShaoZWWXCJ22}}
& \revised{\underline{1.514}} & \revised{\underline{3.500}} & \revised{\underline{3.41\%}}
& \revised{\textbf{1.256}} & \revised{\textbf{2.636}} & \revised{\underline{2.63\%}}
& \revised{\underline{1.569}} & \revised{\underline{3.582}} & \revised{\underline{3.55\%}}
& \revised{1.861} & \revised{\underline{4.312}} & \revised{4.37\%} \\

TESTAM~\citep{lee2024testam}
& \revised{1.530} & \revised{3.566} & \revised{3.44\%}
& \revised{1.297} & 2.781 & 2.74\%
& \revised{1.593} & \revised{3.681} & \revised{3.59\%}
& \underline{1.849} & \revised{4.320} & \underline{4.34\%} \\

AutoSTF~\citep{lyu2025autostf}
& 1.548 & 3.599 & 3.44\%
& 1.298 & \revised{2.758} & 2.72\%
& 1.613 & 3.712 & 3.61\%
& 1.881 & 4.389 & 4.36\% \\

ST-SSDL~\citep{gao2025stssdl}
& 1.579 & 3.678 & 3.51\%
& 1.304 & 2.765 & \revised{2.70\%}
& 1.642 & 3.762 & 3.65\%
& 1.946 & 4.555 & 4.55\% \\

MSTHG~\citep{msthg2026}
& 1.586 & 3.683 & 3.59\%
& 1.323 & 2.816 & 2.82\%
& 1.651 & 3.801 & 3.75\%
& 1.935 & 4.490 & 4.56\% \\

RSTIB-MLP~\citep{rstib2025}
& 1.600 & 3.716 & 3.61\%
& 1.348 & 2.870 & 2.84\%
& 1.665 & 3.832 & 3.77\%
& 1.940 & 4.532 & 4.60\% \\

FaST~\citep{fast2026}
& 1.720 & 4.153 & 3.81\%
& 1.366 & 2.983 & 2.86\%
& 1.762 & 4.147 & 3.92\%
& 2.232 & 5.368 & 5.13\% \\

HyperD~\citep{shao2026hyperd}
& 1.796 & 4.275 & 4.45\%
& 1.643 & 3.889 & 4.07\%
& 1.856 & 4.363 & 4.62\%
& 2.109 & 4.948 & 5.24\% \\

\midrule
\rowcolor{candorrow}
CANDOR (Ours)
& \textbf{1.508} & \textbf{3.500} & \textbf{3.34\%}
& \revised{\underline{1.257}} & \revised{\underline{2.647}} & \textbf{2.59\%}
& \textbf{1.565} & \textbf{3.582} & \textbf{3.46\%}
& \textbf{1.846} & \textbf{4.308} & \textbf{4.31\%} \\

\bottomrule
\end{tabular}
}
}
\end{table}

\textbf{Traffic forecasting.}
{\color{black}Table~\ref{tab:traffic_comparison} shows that CANDOR achieves the lowest
overall MAE, RMSE, and MAPE on both traffic datasets. On METR-LA, it
ranks first at all reported horizons, reducing Overall MAE from $2.870$
to $2.854$ ($0.56\%$) and Overall RMSE from $5.879$ to $5.860$
($0.33\%$) relative to D$^2$STGNN, the strongest baseline. The MAE
improvement increases to $1.29\%$ at 60 min, indicating sustained gains
at longer horizons. On PEMS-BAY, where competing methods are more
tightly clustered, CANDOR reduces overall MAE and MAPE by $0.39\%$ and
$2.00\%$, respectively, while overall RMSE is nearly tied with
D$^2$STGNN. D$^2$STGNN achieves slightly lower MAE and RMSE at 15 min,
whereas CANDOR ranks first on all three metrics at 30 and 60 min.
These results support the effectiveness of CANDOR across heterogeneous
congestion dynamics and smoother traffic regimes, with gains varying
across forecasting horizons and traffic conditions.}

\newcommand{\graycell}[1]{\cellcolor{gray!8}#1}

\begin{table}[htbp]
\centering
\vspace{-10pt}
\caption{
Mean long-horizon forecasting performance over three seeds. The same eight baselines are selected from 16
candidates according to their MSE averaged across all three basins and all reported forecasting horizons.
Complete results for all methods and horizons are provided in Appendix~\ref{app:res}.
}

\label{tab:water_quality_long_horizon}

\vspace{4pt}

\scriptsize
\setlength{\tabcolsep}{1.25pt}
\renewcommand{\arraystretch}{0.90}

\resizebox{\textwidth}{!}{
\begin{tabular}{@{}cc|cc|cc|cc|cc|cc|cc|cc|cc|cc@{}}
\toprule

\multicolumn{2}{c|}{\textbf{Model}}
& \multicolumn{2}{c|}{\textbf{FaST}}
& \multicolumn{2}{c|}{\textbf{MSTHG}}
& \multicolumn{2}{c|}{\textbf{TimeMixer++}}
& \multicolumn{2}{c|}{\textbf{LatentTSF}}
& \multicolumn{2}{c|}{\textbf{Pathformer}}
& \multicolumn{2}{c|}{\textbf{PhaseFormer}}
& \multicolumn{2}{c|}{\textbf{DTAF}}
& \multicolumn{2}{c|}{\textbf{TimeBridge}}
& \multicolumn{2}{c}{\textbf{CANDOR}} \\

\multicolumn{2}{c|}{\textbf{Year}}
& \multicolumn{2}{c|}{\citeyearpar{fast2026}}
& \multicolumn{2}{c|}{\citeyearpar{msthg2026}}
& \multicolumn{2}{c|}{\citeyearpar{wang2025timemixer++}}
& \multicolumn{2}{c|}{\citeyearpar{latenttsf2026}}
& \multicolumn{2}{c|}{\citeyearpar{pathformer2024}}
& \multicolumn{2}{c|}{\citeyearpar{phaseformer2026}}
& \multicolumn{2}{c|}{\citeyearpar{dtaf2026}}
& \multicolumn{2}{c|}{\citeyearpar{timebridge2025}}
& \multicolumn{2}{c}{\textbf{Ours}} \\

\cmidrule(lr){3-4}
\cmidrule(lr){5-6}
\cmidrule(lr){7-8}
\cmidrule(lr){9-10}
\cmidrule(lr){11-12}
\cmidrule(lr){13-14}
\cmidrule(lr){15-16}
\cmidrule(lr){17-18}
\cmidrule(lr){19-20}

\multicolumn{2}{c|}{\textbf{Metric}}
& \textbf{MSE} & \textbf{MAE}
& \textbf{MSE} & \textbf{MAE}
& \textbf{MSE} & \textbf{MAE}
& \textbf{MSE} & \textbf{MAE}
& \textbf{MSE} & \textbf{MAE}
& \textbf{MSE} & \textbf{MAE}
& \textbf{MSE} & \textbf{MAE}
& \textbf{MSE} & \textbf{MAE}
& \textbf{MSE} & \textbf{MAE} \\

\midrule


\multirow{3}{*}{\rotatebox[origin=c]{90}{\textbf{030501}}}
& 48
& \underline{0.308} & 0.336
& 0.347 & 0.359
& 0.313 & 0.342
& 0.321 & 0.357
& 0.311 & \underline{0.334}
& 0.317 & 0.344
& 0.320 & 0.344
& 0.311 & 0.339
& \graycell{\textbf{0.289}} & \graycell{\textbf{0.328}} \\


& 144
& 0.604 & 0.495
& \underline{0.558} & \underline{0.494}
& 0.611 & 0.502
& 0.613 & 0.523
& 0.608 & 0.495
& 0.619 & 0.507
& 0.623 & 0.505
& 0.618 & 0.502
& \graycell{\textbf{0.544}} & \graycell{\textbf{0.478}} \\


& 240
& 0.802 & 0.589
& \underline{0.706} & \underline{0.570}
& 0.811 & 0.598
& 0.799 & 0.616
& 0.807 & 0.589
& 0.815 & 0.599
& 0.821 & 0.598
& 0.820 & 0.596
& \graycell{\textbf{0.665}} & \graycell{\textbf{0.549}} \\


\midrule


\multirow{3}{*}{\rotatebox[origin=c]{90}{\textbf{031300}}}
& 48
& \underline{0.213} & \underline{0.294}
& 0.227 & 0.310
& 0.217 & 0.299
& 0.248 & 0.332
& 0.220 & 0.298
& 0.231 & 0.312
& 0.237 & 0.314
& 0.222 & 0.301
& \graycell{\textbf{0.205}} & \graycell{\textbf{0.287}} \\


& 144
& \underline{0.441} & \underline{0.454}
& 0.469 & 0.481
& 0.447 & 0.460
& 0.468 & 0.482
& 0.451 & 0.461
& 0.465 & 0.470
& 0.471 & 0.472
& 0.468 & 0.468
& \graycell{\textbf{0.420}} & \graycell{\textbf{0.449}} \\


& 240
& \underline{0.557} & \underline{0.525}
& 0.585 & 0.552
& 0.559 & 0.528
& 0.578 & 0.548
& 0.581 & 0.537
& 0.584 & 0.539
& 0.591 & 0.542
& 0.593 & 0.539
& \graycell{\textbf{0.528}} & \graycell{\textbf{0.520}} \\


\midrule


\multirow{3}{*}{\rotatebox[origin=c]{90}{\textbf{180102}}}
& 48
& \underline{0.725} & 0.441
& 0.748 & 0.460
& 0.726 & 0.443
& 0.739 & 0.473
& 0.732 & \underline{0.440}
& 0.752 & 0.455
& 0.743 & 0.451
& 0.745 & 0.445
& \graycell{\textbf{0.712}} & \graycell{\textbf{0.440}} \\


& 144
& \underline{1.390} & \underline{0.652}
& 1.404 & 0.688
& 1.394 & 0.659
& 1.394 & 0.701
& 1.409 & 0.653
& 1.440 & 0.676
& 1.429 & 0.672
& 1.418 & 0.660
& \graycell{\textbf{1.302}} & \graycell{\textbf{0.644}} \\


& 240
& 1.879 & \underline{0.789}
& 1.912 & 0.835
& 1.901 & 0.800
& \underline{1.846} & 0.835
& 1.890 & 0.791
& 1.931 & 0.814
& 1.924 & 0.810
& 1.950 & 0.803
& \graycell{\textbf{1.763}} & \graycell{\textbf{0.780}} \\


\bottomrule
\end{tabular}
}

\end{table}

\textbf{Water-quality forecasting.}
Table~\ref{tab:water_quality_long_horizon} reports long-horizon
forecasting results on sparse directed river networks. CANDOR achieves
the lowest MSE averaged over all five forecasting horizons on each
basin, reducing MSE by $5.47\%$, $4.68\%$, and $5.29\%$ relative to
the strongest baseline on BASIN030501, BASIN031300, and BASIN180102,
respectively.
Unlike traffic forecasting, generic spatiotemporal models do not consistently dominate temporal forecasters on these datasets. This suggests that sparse topology and strong node-specific temporal regularities can limit uniform spatial aggregation. CANDOR addresses this regime by decomposing local dynamics
before dependency learning, modeling delayed interactions over directed physical support, and complementing them with functional relations
beyond topology. The resulting advantage across basins and horizons shows that CANDOR can exploit spatial propagation without assuming that it is uniformly informative. Across the three seeds, CANDOR's worst-seed MSE is lower than the corresponding baseline's best-seed MSE in 235 of 240 basin-horizon-baseline comparisons across all 16 baselines (Appendix~\ref{app:res}).

\textbf{{\color{black}Cross-system comparison.}}
{\color{black}Baseline rankings reveal different modeling priorities across systems.
D$^2$STGNN is the strongest traffic baseline but is less competitive
on river datasets, particularly at longer horizons, where several
temporal forecasters achieve lower errors. This contrast suggests
that propagation mechanisms effective for traffic do not transfer
uniformly across systems. CANDOR performs consistently well in both
domains by combining directed physical propagation with
background-conditioned functional relations. For an aggregate comparison, we normalize each method's primary error by the minimum on each dataset and average across all five
datasets with equal weights, using unrounded three-seed means.
The primary metric is Overall MAE for traffic and five-horizon
average MSE for water quality. CANDOR ranks first, reducing mean
normalized error by $7.25\%$ relative to the runner-up MSTHG.}

\subsection{Ablation Studies}

\begin{wrapfigure}{r}{0.52\textwidth}
\vspace{-3.5em}
\centering
\includegraphics[width=\linewidth]{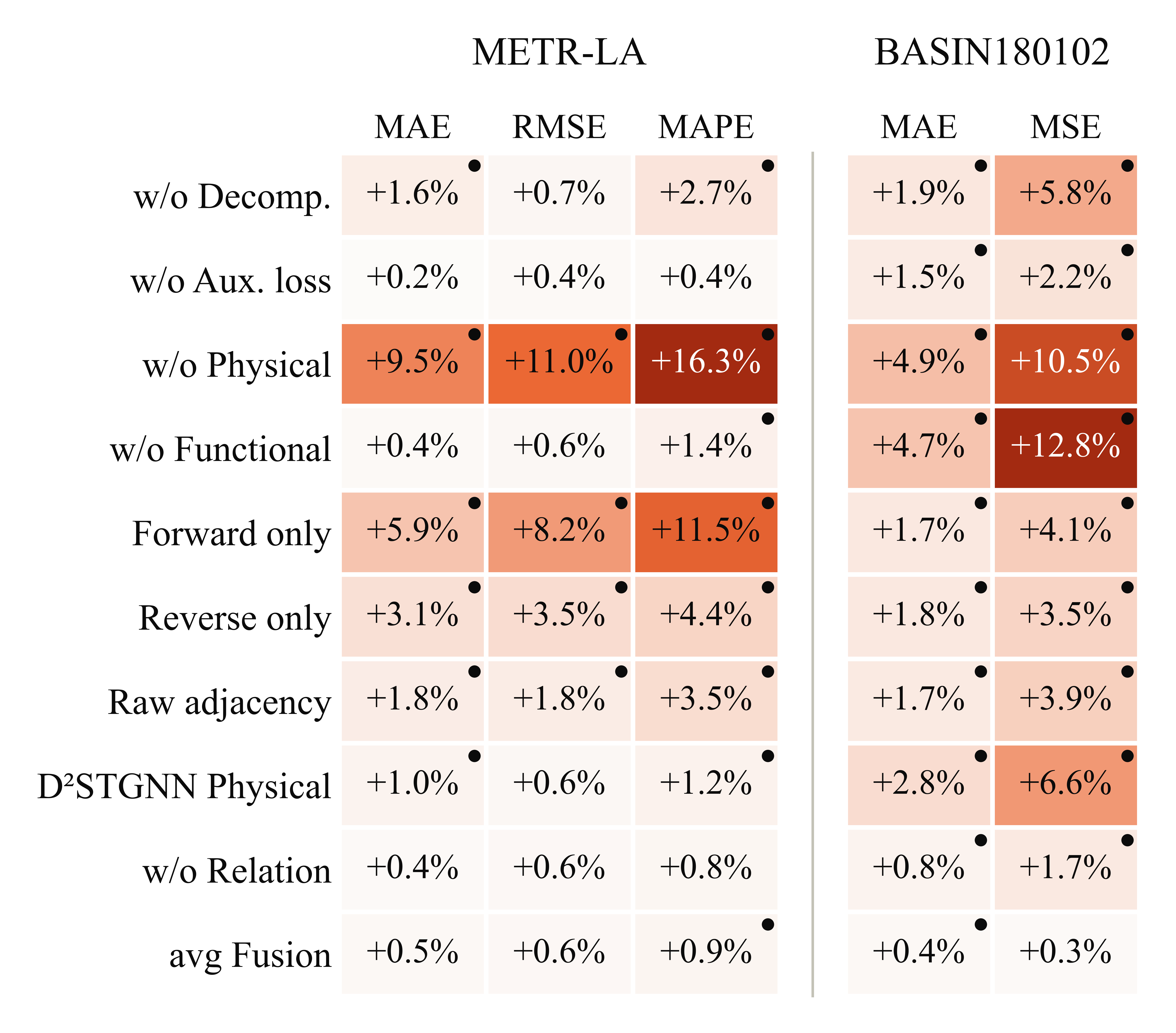}
\vspace{-2em}   
\caption{Seed-paired relative degradation (\%) under each ablation variant. Dots denote changes exceeding twice the full-model seed variation.}
\label{fig:ablation_main}
\vspace{-0em}
\end{wrapfigure}

{\color{black}Figure~\ref{fig:ablation_main} evaluates the modeling chain on two representative datasets; results on all five
datasets are reported in Appendix~\ref{abal}. The ``w/o Decomp.''
variant removes the decomposition and passes patch
representations directly to both relational branches, whereas ``w/o Aux. loss''
retains the components but removes their reconstruction and
role-consistency objectives. ``w/o Physical'' and ``w/o Functional''
remove the two relational branches, while ``Forward only'' and
``Reverse only'' retain one physical direction. ``Raw adjacency''
replaces the learned physical operator with normalized topology.
\revised{``D$^2$STGNN Physical'' replaces CANDOR's entire physical module
with a D$^2$STGNN decoupling layer~\citep{ShaoZWWXCJ22}, retaining
the same patch grid, upstream decomposition, and functional branch.
Its combined diffusion and inherent forecast is adaptively fused
with the functional forecast.}
Finally, ``w/o Relation'' removes functional relation modeling, while
``avg Fusion'' replaces adaptive fusion with uniform averaging. 

The results support the progression from decomposition to relational
learning. Removing decomposition increases METR-LA MAE by $1.6\%$ and
BASIN180102 MSE by $5.8\%$; removing only its auxiliary objectives
increases BASIN180102 MAE/MSE by $1.5\%/2.2\%$. This shows that the
benefit arises from both component representations and their assigned
roles. The relational branches are complementary across systems.
Removing physical propagation increases METR-LA MAE/MAPE by
$9.5\%/16.3\%$, whereas removing the physical and functional branches
on BASIN180102 increases MSE by $10.5\%$ and $12.8\%$, respectively.
Both single-direction variants degrade performance. Replacing the
learned operator with raw adjacency further increases METR-LA MAPE by
$3.5\%$ and BASIN180102 MSE by $3.9\%$, confirming that topology alone
does not capture direction, delay, and history-dependent propagation.
\revised{The D$^2$STGNN replacement increases METR-LA MAE by $1.0\%$
and BASIN180102 MAE/MSE by $2.8\%/6.6\%$.}
Relation modeling and adaptive fusion contribute up to $3.4\%$ and
$2.4\%$ in MSE in the complete matrix. Together with the role alignments in
Figure~\ref{fig:mechanism_compact}, these results support CANDOR's
central premise: it organizes local dynamics into evolution-specific roles, enabling ordered physical propagation and complementary functional relations to be learned and adaptively combined.}

\subsection{Model Analysis}
\begin{figure*}[htbp]
\vspace{-0.3em}
\centering
\includegraphics[width=1\textwidth]{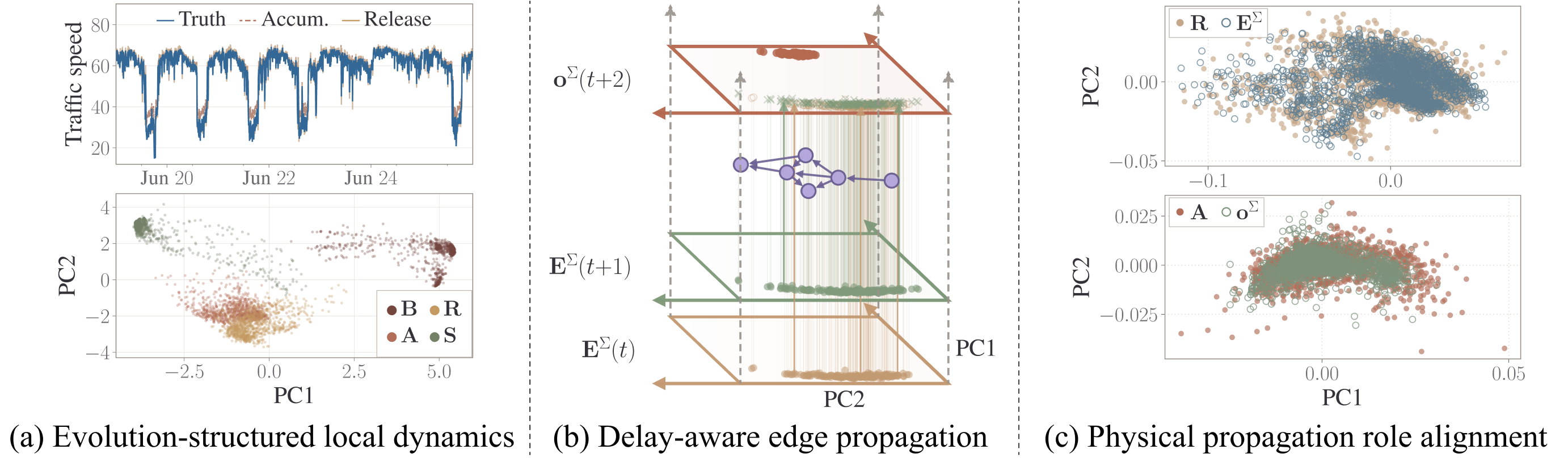}
\caption{Decomposition and physical propagation on METR-LA.
(a) Reference-consistent reconstruction and embeddings of the four
components. (b) Delayed mapping from release-associated residuals
$\mathbf{E}^{\Sigma}$ to incoming graph contribution
$\mathbf{o}^{\Sigma}$. (c) Role alignment of $\mathbf{R}$ with
$\mathbf{E}^{\Sigma}$ and $\mathbf{A}$ with $\mathbf{o}^{\Sigma}$.}
\label{fig:mechanism_compact}
\vspace{-0.4em}
\end{figure*}

\begin{figure*}[htbp]
\centering
\includegraphics[width=0.95\textwidth]{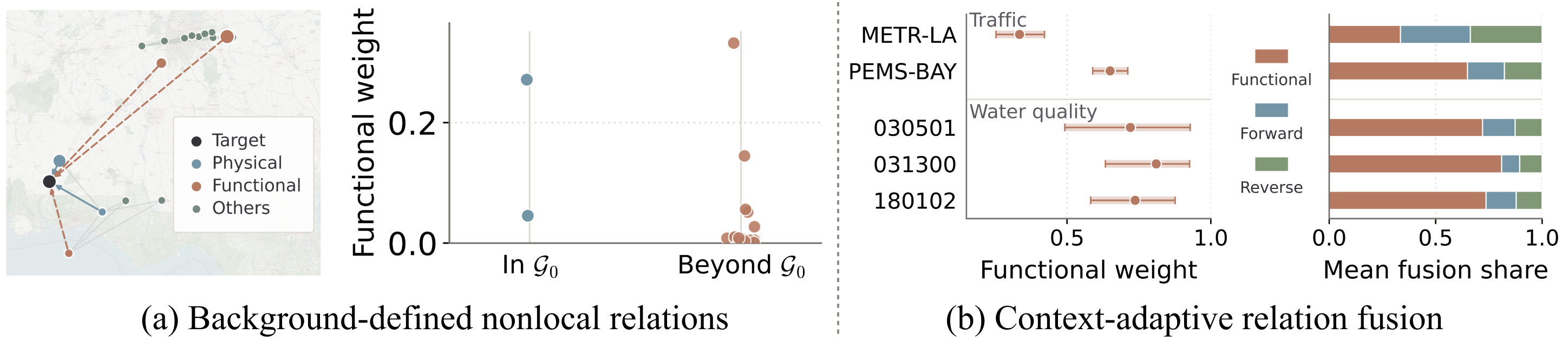}
\caption{Functional relations beyond topology and context-adaptive fusion.
(a) Background dynamics identify nonlocal functional relations beyond
physical adjacency. (b) CANDOR adjusts the use of functional, forward,
and reverse-support forecasts across systems.}
\label{fig:system_analysis}
\vspace{-0.4em}
\end{figure*}

Figure~\ref{fig:mechanism_compact} validates the bridge from
decomposition to physical propagation. Panel (a) shows that the
reference-consistent reconstructions preserve the trajectory while
background, accumulation, release, and shock form distinct embedding
regions, demonstrating information retention and role differentiation.
Let $\mathbf{E}^{\Sigma}$ and $\mathbf{o}^{\Sigma}$ denote the combined
propagation residual and incoming graph contribution. Panel (b) maps
$\mathbf{E}^{\Sigma}(t)$ and $\mathbf{E}^{\Sigma}(t+1)$ to the later
$\mathbf{o}^{\Sigma}(t+1)$, while panel (c) aligns $\mathbf{R}$ with
$\mathbf{E}^{\Sigma}$ and $\mathbf{A}$ with $\mathbf{o}^{\Sigma}$.
These results support delayed propagation from release to subsequent accumulation. Figure~\ref{fig:system_analysis} examines their complementarity with
physical propagation. {\color{black}Panel (a) shows that background-conditioned
functional relations extend beyond $\mathcal{G}_0$, capturing persistent
nonlocal dependencies without transient deviations. Panel (b) shows
context-dependent allocation: METR-LA favors physical propagation,
whereas PEMS-BAY and the water-quality datasets assign more weight to
functional relations. This indicates that fusion preserves distinct
pathways and supplements topology when needed. Further validation and
physical interpretations are provided in
Appendix~\ref{app:mechanism_analysis}.}

\section{Conclusion}

{\color{black}We introduce \textbf{CANDOR}, a component-aware framework that
decomposes local dynamics before learning node dependencies.
Separating persistent, evolving, and abrupt dynamics grounds
delayed physical propagation and nonlocal functional dependence
in distinct dynamical roles. Across human-driven traffic and
natural river systems, CANDOR achieves the lowest aggregate
normalized error, outperforming the overall runner-up by $7.25\%$.
Water-quality results highlight the limitations of spatial modeling
under sparse topology and heterogeneous delays, motivating
complementary relational mechanisms. These findings support
decomposition before dependency learning as an effective principle
for spatiotemporal representation learning.}





\bibliography{iclr2027_conference}
\bibliographystyle{iclr2027_conference}

\appendix
\section{Appendix}

{\color{black}
\subsection{Physical Motivation and Reference-State Interpretation}
\label{app:decomposition_physics}

We explain the mechanisms that motivate separating local dynamics
and how they inform CANDOR's two background references. The physical
models provide a basis for the intended dynamical roles and CANDOR
learns their representations from observed trajectories.

\paragraph{Mechanistic basis.}
In traffic, the cell transmission model describes vehicle storage
through an inflow-outflow balance \citep{daganzo1995cell}:
\begin{equation}
n_i(t+h)-n_i(t)
=
\int_t^{t+h}
\left(q_i^{\mathrm{in}}(s)-q_i^{\mathrm{out}}(s)\right)\,ds,
\label{eq:app_traffic_balance}
\end{equation}
where $n_i$ is the number of vehicles in road cell $i$, and
$q_i^{\mathrm{in}}$ and $q_i^{\mathrm{out}}$ are flow rates.
Queues grow when inflow exceeds discharge and can dissipate when
this imbalance reverses. Changes in demand or capacity alter this
evolution, while incidents can introduce abrupt perturbations.
For speed forecasting, these processes appear through departures
from prevailing traffic conditions and subsequent recovery.
Macroscopic speed models additionally describe relaxation toward
a traffic-dependent reference velocity alongside spatial
interactions \citep{treiber1999derivation}.

In rive network, dissolved oxygen similarly reflects competing local processes
and transport. A schematic concentration balance is
\begin{equation}
\begin{aligned}
\frac{dC_i}{dt}
={}&\mathcal{T}_i(\mathbf{C})
+k_{a,i}(C_{\mathrm{sat},i}-C_i)\\
&+p_{\mathrm{photo},i}
-r_{\mathrm{resp},i}
-r_{\mathrm{ox},i}
+\ell_{\mathrm{ext},i},
\end{aligned}
\label{eq:app_oxygen_balance}
\end{equation}
where $\mathcal{T}_i$ represents net advective and dispersive
transport, $k_{a,i}$ controls air-water exchange, and
$C_{\mathrm{sat},i}$ is the saturation concentration.
The remaining terms represent photosynthetic production,
respiration, oxidative demand, and external loading.
Their relative strengths vary with environmental conditions,
producing oxygen depletion and recovery as well as recurring
temporal patterns \citep{sullivan2011modeling}.
Thus, similar concentration levels can occur during different
stages of evolution. These mechanisms motivate distinguishing persistent conditions, the formation and relaxation
of local deviations, and abrupt disturbances.

\paragraph{Why use current and subsequent references?}
Both systems contain responses defined relative to changing
operating conditions. Traffic models use a condition-dependent
reference velocity, while the oxygen exchange term depends on
the difference from saturation. These physical references motivate
learning an effective background when the governing variables and
parameters are unavailable. The effect of reference evolution can be illustrated by a local
relaxation model:
\begin{equation}
\dot{x}_i(t)
=
-\kappa_i(t)\bigl(x_i(t)-b_i(t)\bigr)
+f_i(t)+\xi_i(t),
\label{eq:app_reference_response}
\end{equation}
where $b_i(t)$ is an evolving reference,
$\kappa_i(t)>0$ is a relaxation coefficient, $f_i(t)$ collects
continuous driving and transport effects, and $\xi_i(t)$ represents
abrupt forcing. Defining the deviation
$r_i(t)=x_i(t)-b_i(t)$ gives
\begin{equation}
\dot{r}_i(t)
=
-\kappa_i(t)r_i(t)
+f_i(t)+\xi_i(t)-\dot{b}_i(t).
\label{eq:app_reference_drift}
\end{equation}
The reference drift contributes explicitly to deviation
dynamics. A change in departure can reflect both state evolution
and a change in the operating baseline. CANDOR encodes this distinction through
Eq.~\ref{eq:evolution_references}. For $p<P$,
\begin{equation}
\mathbf{V}_{i,p}
=
\mathbf{U}_{i,p}
-
\left(\mathbf{B}_{i,p+1}-\mathbf{B}_{i,p}\right).
\label{eq:app_reference_shift}
\end{equation}
The current-reference view describes the departure under present
conditions. The shifted-reference view expresses the same state
under the subsequent observed baseline, explicitly incorporating
one-step reference evolution. Both references remain within the
observed history. This provides complementary anchors for
accumulation and release, while temporal differences and
context-dependent gates guide their extraction according to
the direction and variation of local changes. These roles concern
departures from a baseline: congestion buildup, for example, can
appear as decreasing speed. Furthermore, the same abrupt disturbance can affect both reference views,
motivating the shared shock component. The two reconstruction
objectives further connect accumulation and release through
background evolution. When reconstruction errors are small,
\begin{equation}
\widetilde{\mathbf{A}}_{i,p}
-\widetilde{\mathbf{R}}_{i,p}
\approx
\widetilde{\mathbf{B}}_{i,p+1}
-\widetilde{\mathbf{B}}_{i,p},
\label{eq:app_reference_consistency}
\end{equation}
where tildes denote outputs before normalization.
The common shock contribution cancels when comparing the two
reconstructions. Role-consistency supervision then associates
release with propagation residuals and subsequent accumulation
with incoming graph contributions. 

These constraints connect
reference-dependent local dynamics to their use in propagation.
The components represent learned dynamical roles and the next background serves as an observed reference.
}

\subsection{Evolution-Structured Dynamics Decomposition}
\label{app:decomposition}

We detail the computations in Section~\ref{sec:decomposition},
omitting the batch dimension.

\paragraph{Background and residual construction.}
We divide each node's history into patches of length $\ell$
and stride $s$, giving $P=\lfloor(L-\ell)/s\rfloor+1$.
For the observation vector $\mathbf{x}_{i,p}\in\mathbb{R}^{\ell}$
of node $i$ at patch $p$, the embedding and background are
\begin{equation}
\mathbf{Z}_{i,p}
=
\operatorname{LN}_{Z}
(\mathbf{W}_{Z}\mathbf{x}_{i,p}+\mathbf{b}_{Z}),
\qquad
\widetilde{\mathbf{B}}
=
\mathcal{C}_{B}(\mathbf{Z}),
\qquad
\mathbf{B}
=
\operatorname{LN}_{B}(\widetilde{\mathbf{B}}),
\label{eq:app_embedding_background}
\end{equation}
where $\mathbf{W}_{Z}\in\mathbb{R}^{D\times\ell}$ and
$\mathbf{b}_{Z}\in\mathbb{R}^{D}$ are learned parameters.
The operator $\mathcal{C}_{B}$ is a depthwise temporal
convolution with kernel size $K$ and $K-1$ zeros padded
on the left.

Equation~\ref{eq:evolution_references} constructs
$\mathbf{U}$ and $\mathbf{V}$ from $\mathbf{B}$.
The shifted reference repeats the final background at the
boundary and accesses only observed history. Thus, although
the convolution is causal, the complete decomposition uses
context across the observed window.

\paragraph{Evolution-dependent component extraction.}
For either residual view $\mathbf{Q}\in\{\mathbf{U},\mathbf{V}\}$,
temporal differences are defined recursively:
\begin{equation}
(\Delta\mathbf{Q})_{i,p}
=
\begin{cases}
\mathbf{0}, & p=1,\\
\mathbf{Q}_{i,p}-\mathbf{Q}_{i,p-1}, & p>1,
\end{cases}
\qquad
\Delta^{2}\mathbf{Q}=\Delta(\Delta\mathbf{Q}).
\label{eq:app_temporal_differences}
\end{equation}
Let $\mathbf{e}_{i}\in\mathbb{R}^{D}$ be a learned node
embedding and $\mathbf{m}_{p}\in\mathbb{R}^{D}$ the calendar
context at the patch start. The latter is obtained by an MLP
over concatenated time-of-day and day-of-week embeddings,
using 288 and seven categories, respectively.
With an affine node projection $\mathcal{P}_{E}$,
the gate input is
\begin{equation}
\boldsymbol{\xi}_{i,p}(\mathbf{Q})
=
\left[
\mathbf{Q}_{i,p}
\Vert(\Delta\mathbf{Q})_{i,p}
\Vert-(\Delta\mathbf{Q})_{i,p}
\Vert(\Delta^{2}\mathbf{Q})_{i,p}
\Vert\mathcal{P}_{E}(\mathbf{e}_{i})
\Vert\mathbf{m}_{p}
\right],
\label{eq:app_gate_inputs}
\end{equation}
where $\Vert$ denotes feature concatenation.
Separate MLPs $f_A$ and $f_R$ produce feature-wise
sigmoid gates:
\begin{equation}
\mathbf{G}_{A,i,p}
=
\sigma(f_A(\boldsymbol{\xi}_{i,p}(\mathbf{U}))),
\qquad
\mathbf{G}_{R,i,p}
=
\sigma(f_R(\boldsymbol{\xi}_{i,p}(\mathbf{V}))).
\end{equation}
The gates weight the original residuals before temporal
convolution:
\begin{equation}
\widetilde{\mathbf{A}}=
\mathcal{C}_{A}(\mathbf{G}_{A}\odot\mathbf{U}),
\qquad
\mathbf{A}=
\operatorname{LN}_{A}(\widetilde{\mathbf{A}}), \qquad
\widetilde{\mathbf{R}}=
\mathcal{C}_{R}(\mathbf{G}_{R}\odot\mathbf{V}),
\qquad
\mathbf{R}=
\operatorname{LN}_{R}(\widetilde{\mathbf{R}}),
\label{eq:app_accumulation_release}
\end{equation}
where $\odot$ denotes element-wise multiplication.
The convolutions $\mathcal{C}_{A}$ and $\mathcal{C}_{R}$
use the same configuration as $\mathcal{C}_{B}$
with independent parameters.

Shock uses separate scalar gates for the two views:
\begin{equation}
g^{S}_{Q,i,p}
=
\sigma\!\left(
f_{S,Q}\!\left(
\left[
|\mathbf{Q}_{i,p}|
\Vert|(\Delta\mathbf{Q})_{i,p}|
\Vert|(\Delta^{2}\mathbf{Q})_{i,p}|
\right]\right)
+
c_Q(\mathbf{e}_{i},\mathbf{m}_{p})
\right),
\label{eq:app_shock_gates}
\end{equation}
where $f_{S,Q}$ scores residual magnitudes and $c_Q$
provides a learned contextual correction.
Each $c_Q$ separately projects the node and calendar
embeddings, concatenates them, and applies an MLP followed
by a scalar projection initialized to zero.
The gate MLPs use hidden width $2D$, GELU activation,
and dropout. Collecting the scalar gates into $\mathbf{g}^{S}_{U}$ and $\mathbf{g}^{S}_{V}$, shock is
\begin{equation}
\widetilde{\mathbf{S}}
=
\mathbf{g}^{S}_{U}\odot\mathcal{P}_{U}(\mathbf{U})
+
\mathbf{g}^{S}_{V}\odot\mathcal{P}_{V}(\mathbf{V}),
\qquad
\mathbf{S}
=
\operatorname{LN}_{S}(\widetilde{\mathbf{S}}),
\label{eq:app_shock_component}
\end{equation}
where $\mathcal{P}_{U}$ and $\mathcal{P}_{V}$ are independent
affine feature projections. Scalar gates are broadcast
across features and need not sum to one. Magnitudes determine
selection, while the projected residuals retain signed information.

\paragraph{Historical integration and temporal projection.}
An affine map $\mathcal{P}_{H}:\mathbb{R}^{4D}\rightarrow
\mathbb{R}^{D}$ integrates the four components through a
residual connection:
\begin{equation}
\mathbf{H}^{(0)}
=
\operatorname{LN}_{H}\!\left(
\mathbf{Z}
+
\alpha\mathcal{P}_{H}
([\mathbf{B}\Vert\mathbf{A}\Vert\mathbf{R}\Vert\mathbf{S}])
\right),
\label{eq:app_history_integration}
\end{equation}
where $\alpha$ is a learned scalar initialized to $0.1$
and $\mathbf{H}^{(0)}\in\mathbb{R}^{N\times P\times D}$.
The background is also retained separately for functional
relation learning. Each component has independent affine temporal and feature
projections. For label $c\in\{B,A,R,S\}$, let
$\mathbf{C}^{B}=\mathbf{B}$,
$\mathbf{C}^{A}=\mathbf{A}$,
$\mathbf{C}^{R}=\mathbf{R}$, and
$\mathbf{C}^{S}=\mathbf{S}$.
The temporal map $\mathcal{T}_{c}$ projects $P$ historical
positions to $P_{\mathrm{out}}$ target patches, and
$\mathcal{F}_{c}$ maps $D$ features to the forecast dimension
$D_{\mathrm{f}}$. Writing
$\mathbf{F}_{c}=\mathcal{F}_{c}(\mathcal{T}_{c}(\mathbf{C}^{c}))$,
we obtain
\begin{equation}
\mathbf{F}^{(0)}
=
\operatorname{LN}_{F}\!\left(
\mathcal{P}_{F}
([\mathbf{F}_{B}\Vert\mathbf{F}_{A}
\Vert\mathbf{F}_{R}\Vert\mathbf{F}_{S}])
+
\mathbf{M}_{\mathrm{out}}
\right),
\label{eq:app_temporal_projection}
\end{equation}
where $\mathcal{P}_{F}$ maps $4D_{\mathrm{f}}$ features
to $D_{\mathrm{f}}$, giving
$\mathbf{F}^{(0)}\in
\mathbb{R}^{N\times P_{\mathrm{out}}\times D_{\mathrm{f}}}$.
Unlike $H$, which counts forecasting time steps,
$P_{\mathrm{out}}$ counts target patches.
The calendar representation
$\mathbf{M}_{\mathrm{out}}\in
\mathbb{R}^{P_{\mathrm{out}}\times D_{\mathrm{f}}}$
uses the shared calendar embedding tables and a separate
projection network, and is broadcast across nodes. Background smoothness and shock-gate sparsity are specified
in Appendix~\ref{app:training_objectives}.
The components encode learned evolution roles without
imposing exact additive reconstruction or physical conservation.

\subsection{Decomposition-Conditioned Dual-Graph Learning}
\label{app:dual_graph}

We detail the two relational branches in
Section~\ref{sec:dual_graph}, omitting the batch dimension. Matrix entry
$(i,j)$ represents information received by node $i$ from node $j$.

{\color{black}
\paragraph{Physical propagation graph.}
Let $\mathbf{J}\in\mathbb{R}_{\geq0}^{N\times N}$ denote
the adjacency of $\mathcal{G}_0$, where $J_{ij}>0$ indicates
a physical edge from $i$ to $j$. Unlike $\mathbf{J}$,
propagation operators use the receiving-node convention:
entry $(i,j)$ weights information from node $j$ to node $i$.
The directional supports are
\begin{equation}
\mathbf{T}^{\rightarrow}
=
\operatorname{RowNorm}(\mathbf{J})^{\top},
\qquad
\mathbf{T}^{\leftarrow}
=
\operatorname{RowNorm}(\mathbf{J}^{\top})^{\top},
\label{eq:app_directional_support}
\end{equation}
where $\operatorname{RowNorm}$ preserves zero rows. For direction $d$, let
$\mathbf{e}_{i,\mathrm{rec}}^{d}$ and
$\mathbf{e}_{i,\mathrm{src}}^{d}\in\mathbb{R}^{D_e}$
denote receiving and source embeddings.
Two learned embedding tables supply these roles, which
are exchanged between directions.
A shared query projection $\mathbf{W}_{q}$ and
offset-specific key projections $\mathbf{W}_{k,\tau}$
define
\begin{equation}
\omega_{ij}^{d,\tau}
=
\operatorname{softmax}_{j}\!\left(
\frac{
(\mathbf{W}_{q}\mathbf{e}_{i,\mathrm{rec}}^{d})^{\top}
(\mathbf{W}_{k,\tau}\mathbf{e}_{j,\mathrm{src}}^{d})
}{\sqrt{D_e}}
\right),
\qquad \tau=1,\ldots,K_t.
\label{eq:app_delay_affinity}
\end{equation}
These affinities vary across edges and temporal positions
but are shared across samples.

Sample-dependent affinities use a descriptor
$\mathbf{z}_{i}^{d}$ formed by concatenating an MLP encoding
of $\mathbf{X}_{t-L+1:t,i}$, calendar embeddings at time $t$,
and $\mathbf{e}_{i,\mathrm{src}}^{d}$.
Learned query and key maps $\mathcal{Q}_{a}$ and
$\mathcal{K}_{a}$ produce $D$-dimensional features:
\begin{equation}
a_{ij}^{d}
=
\operatorname{softmax}_{j}\!\left(
\frac{
\mathcal{Q}_{a}(\mathbf{z}_{i}^{d})^{\top}
\mathcal{K}_{a}(\mathbf{z}_{j}^{d})
}{\sqrt{D}}
\right).
\label{eq:app_sample_affinity}
\end{equation}
The affinities $\mathbf{a}^{d}$ are shared across temporal
positions within each sample.
Equation~\ref{eq:physical_graph} combines them with
$\mathbf{T}^{d}$ and $\boldsymbol{\omega}^{d,\tau}$.
Normalization is performed over source nodes separately
for each direction and offset, with a small positive
denominator floor for numerical stability.
The resulting operators are computed once per observed
window and reused during forecasting.

To construct propagation states, each directional path
applies a scalar node-calendar gate to $\mathbf{H}^{(0)}$,
giving $\widetilde{\mathbf{H}}^{d}$.
A GRU followed by causal attention with positional encoding
produces the temporal representation $\mathbf{L}^{d}$.
A learned backcast map $\mathcal{P}_{\mathrm{back}}^{d}$
then defines the propagation residual
\begin{equation}
\mathbf{E}^{d}
=
\operatorname{LN}_{d}\!\left(
\widetilde{\mathbf{H}}^{d}
-
\operatorname{ReLU}
(\mathcal{P}_{\mathrm{back}}^{d}(\mathbf{L}^{d}))
\right).
\label{eq:app_propagation_residual}
\end{equation}

{\color{black}
To account for abrupt local disturbances, we further
modulate the propagation residuals using the shock
component obtained from the decomposition module.
Let $\widetilde{\mathbf{S}}$ denote the pre-normalization
shock representation in Eq.~\ref{eq:app_shock_component}.
The available shock context is defined as
\begin{equation}
\mathbf{S}_{i,p}^{\mathrm{ctx}}=
\begin{cases}
\widetilde{\mathbf{S}}_{i,p}, & p\le P,\\[3pt]
\displaystyle\frac{1}{w_p}
\sum_{r=p-w_p}^{p-1}\mathbf{S}_{i,r}^{\mathrm{ctx}},
& p>P,
\end{cases}
\quad w_p=\min(K_t,p-1).
\label{eq:app_shock_context}
\end{equation}
For future patches, the shock context is recursively
estimated from previously available contexts without
accessing future ground-truth observations.

We quantify the local shock intensity as
\begin{equation}
s_{i,p}=1-\exp\left(
-\frac{\|\mathbf{S}_{i,p}^{\mathrm{ctx}}\|_1}{D}
\right).
\label{eq:app_shock_strength}
\end{equation}
The shock representation is combined with node
and calendar context:
\begin{equation}
\mathbf{z}_{i,p}^{S}
=f_S\left(
[\operatorname{LN}_0(\mathbf{S}_{i,p}^{\mathrm{ctx}})
\Vert\mathbf{e}_i\Vert\mathbf{c}_p^S]
\right),
\label{eq:app_shock_encoding}
\end{equation}
where $f_S$ is a shared MLP, $\operatorname{LN}_0$ denotes layer normalization
without learnable affine parameters, $\mathbf{e}_i$
is the shared node embedding, and $\mathbf{c}_p^S$
is the shared calendar context constructed from
the forward path's gate calendar embeddings.

Direction-specific modulation parameters are given by
\begin{equation}
\boldsymbol{\gamma}_{i,p}^{d}
=\tanh(\mathcal{P}_{\gamma}^{d}(\mathbf{z}_{i,p}^{S})),
\quad
\boldsymbol{\beta}_{i,p}^{d}
=\tanh(\mathcal{P}_{\beta}^{d}(\mathbf{z}_{i,p}^{S})),
\label{eq:app_shock_parameters}
\end{equation}
where $\mathcal{P}_{\gamma}^{d}$ and $\mathcal{P}_{\beta}^{d}$
are direction-specific affine maps producing
$D$-dimensional modulation vectors.

The resulting shock-modulated propagation state is
\begin{equation}
\begin{aligned}
\mathbf{E}_{\mathrm{used},i,p}^{d}
={}&\mathbf{E}_{i,p}^{d}+\alpha_S s_{i,p}
\left(
\boldsymbol{\gamma}_{i,p}^{d}
\odot\operatorname{LN}_0(\mathbf{E}_{i,p}^{d})
+\boldsymbol{\beta}_{i,p}^{d}
\right),
\end{aligned}
\label{eq:app_shock_modulation}
\end{equation}
where $\alpha_S$ controls the modulation scale.
The additive term corresponds to
$\Delta_S^d$ in Eq.~\ref{eq:physical_states}.
Shock modulation adjusts the propagated features.
The unmodulated residual $\mathbf{E}^d$ is retained
for role-consistency supervision, while
$\mathbf{E}_{\mathrm{used}}^d$ is used for
spatial aggregation.
}

For spatial order $m=1,\ldots,K_s$, where $K_s$ is the
maximum spatial order, we obtain
$\mathbf{W}^{d,\tau}_{[m]}$ by taking the matrix power
$(\mathbf{W}^{d,\tau})^{m}$ and setting its diagonal to zero.
Powers are computed from the original operator before
diagonal removal, without subsequent renormalization.

{\color{black}
For a window ending at patch $p$, the latest $K_t$
shock-modulated propagation states from
$\mathbf{E}_{\mathrm{used}}^{d}$ are concatenated
along the feature dimension, transformed by a learned
linear map and ReLU, and reshaped into
$\boldsymbol{\Psi}_{j,p,\tau}^{d}\in\mathbb{R}^{D}$.
}
Temporal positions are ordered from oldest to newest and short histories are padded by repeating their earliest state.
The transformation mixes information within the window,
so $\tau$ indexes a transformed temporal position. The aggregation at spatial order $m$ is
\begin{equation}
\mathbf{g}_{i,p}^{d,[m]}
=
\sum_{\tau=1}^{K_t}\sum_{j=1}^{N}
[\mathbf{W}^{d,\tau}_{[m]}]_{ij}
\boldsymbol{\Psi}_{j,p,\tau}^{d},
\label{eq:app_multihop_aggregation}
\end{equation}
where the first-order case matches
Eq.~\ref{eq:physical_graph}.
The complete graph output combines all spatial orders
with a local temporal summary:
\begin{equation}
\mathbf{o}_{i,p}^{d}
=
\mathcal{P}_{g}^{d}\!\left(
\left[
\frac{1}{K_t}\sum_{\tau=1}^{K_t}\boldsymbol{\Psi}_{i,p,\tau}^{d}
\;\middle\Vert\;
\mathbf{g}_{i,p}^{d,[1]}
\Vert\cdots\Vert
\mathbf{g}_{i,p}^{d,[K_s]}
\right]\right),
\label{eq:app_localized_propagation}
\end{equation}
where $\mathcal{P}_{g}^{d}$ is an affine projection followed
by dropout. The local summary retains same-node information,
while the graph terms exclude diagonal contributions.


{\color{black}
At each future patch, the forward-propagation and
reverse-support paths separately predict the next
latent state by combining their temporal states
with the corresponding graph contributions.
For $p=P,\ldots,P+P_{\mathrm{out}}-1$, we compute
\begin{equation}
\widehat{\mathbf{h}}_{i,p+1}^{d}
=\mathbf{L}_{i,p}^{d}+\mathbf{o}_{i,p}^{d},
\quad
\overline{\mathbf{h}}_{i,p+1}
=\frac{1}{2}\left(
\widehat{\mathbf{h}}_{i,p+1}^{\rightarrow}
+\widehat{\mathbf{h}}_{i,p+1}^{\leftarrow}
\right).
\label{eq:app_coupled_rollout}
\end{equation}
The shared prediction is fed back as the latent
input to both directional paths:
\begin{equation}
\widetilde{\mathbf{h}}_{i,p+1}^{d}
=g_{i,p+1}^{d}\overline{\mathbf{h}}_{i,p+1},
\label{eq:app_coupled_feedback}
\end{equation}
where $g_{i,p+1}^{d}$ is the direction-specific
node-calendar gate.

Each path then independently updates its temporal
memory, extracts a new propagation residual through
temporal backcasting, and applies shock-conditioned
modulation before the next graph propagation step.
Although the feedback state is shared, the temporal
states, propagation features, and graph operators
remain direction-specific throughout forecasting. The directional predictions are retained separately
and projected into the forecast representation space:
\begin{equation}
\mathbf{F}_{i,q}^{d}
=\operatorname{LN}_{F}^{d}\left(
\mathcal{P}_{\mathrm{forecast}}^{d}
(\widehat{\mathbf{h}}_{i,P+q}^{d})
\right),
\quad q=1,\ldots,P_{\mathrm{out}},
\label{eq:app_physical_forecast}
\end{equation}
where $\mathcal{P}_{\mathrm{forecast}}^{d}$ maps
$D$-dimensional latent states to $D_{\mathrm{f}}$ features,
and $\operatorname{LN}_{F}^{d}$ applies layer normalization. This yields $\mathbf{F}^{\rightarrow}$ and
$\mathbf{F}^{\leftarrow}$ on the target-patch grid,
which are preserved as separate inputs to the
context-adaptive fusion module.
Reverse support provides complementary predictive
context without implying reverse physical transport.
}
}

\paragraph{Functional relation graph.}
For each node, we summarize $\mathbf{B}_{i,1:P}$ using its
mean level, first-to-last change, most recent increment,
mean increment, and the final state of a GRU over the
background trajectory. An MLP and layer normalization map
these summaries to a base descriptor
$\overline{\mathbf{h}}_{i}^{B}\in\mathbb{R}^{D_b}$,
where $D_b$ is the relation descriptor dimension.
A separate MLP and layer normalization map the three change
summaries to $\mathbf{v}_{i}^{B}\in\mathbb{R}^{D_b}$.
Change summaries are zero when $P=1$. Node and calendar context generate feature-wise scale
$\boldsymbol{\gamma}_{i}^{B}$ and shift
$\boldsymbol{\beta}_{i}^{B}$, giving
\begin{equation}
\mathbf{h}_{i}^{B}
=
\operatorname{LN}_{B,\mathrm{rel}}\!\left(
(1+\boldsymbol{\gamma}_{i}^{B})
\odot\overline{\mathbf{h}}_{i}^{B}
+
\boldsymbol{\beta}_{i}^{B}
\right).
\label{eq:app_background_context}
\end{equation}
The context combines a projected node embedding with pooled
calendar embeddings from historical and target patches.

Using these descriptors, Equation~\ref{eq:functional_graph}
computes scores for $R_{\mathrm{rel}}$ relation channels,
with $\mathbf{M}_{r},\mathbf{N}_{r}
\in\mathbb{R}^{D_b\times D_b}$.
For receiving node $i$, let $\mathcal{K}_{i}$ contain
the highest-scoring $\min(k,(N-1)R_{\mathrm{rel}})$ pairs
$(j,r)$ with $j\neq i$. The sparse weights are
\begin{equation}
\pi_{ij}^{r}
=
\begin{cases}
\displaystyle
\frac{\exp(s_{ij}^{r})}
{\sum_{(j',r')\in\mathcal{K}_{i}}\exp(s_{ij'}^{r'})},
& (j,r)\in\mathcal{K}_{i},\\[6pt]
0, & \text{otherwise}.
\end{cases}
\label{eq:app_joint_relation_selection}
\end{equation}
An empty candidate set gives zero weights. Selection is joint
over neighbors and relation channels, so multiple channels
may connect the same node pair.

Let $\mathcal{V}_{r}$ denote a relation-specific value map,
implemented by a shared feature reduction, a channel-specific
projection, and a shared restoration to $D$ dimensions.
The historical relational context is
\begin{equation}
\mathbf{C}_{i,p}^{\mathrm{func}}
=
\sum_{j=1}^{N}\sum_{r=1}^{R_{\mathrm{rel}}}
\pi_{ij}^{r}\,
\mathcal{V}_{r}(\mathbf{H}^{(0)}_{j,p}).
\label{eq:app_functional_aggregation}
\end{equation}
Thus, background descriptors determine the weights, whereas
values retain information from all four components.

Independent affine temporal and feature projections followed
by layer normalization map
$\mathbf{C}^{\mathrm{func}}$ to a target-patch context.
This context is combined with $\mathbf{F}^{(0)}$,
$\mathbf{h}^{B}$, node embeddings, and target calendar
features through a shared MLP. Two output heads produce
a forecast correction $\boldsymbol{\delta}_{\mathrm{func}}$
and a sigmoid gate $\boldsymbol{\Lambda}_{\mathrm{func}}$:
\begin{equation}
\mathbf{F}^{\mathrm{func}}
=
\operatorname{LN}_{\mathrm{func}}\!\left(
\mathbf{F}^{(0)}
+
\alpha_{\mathrm{func}}\,
\boldsymbol{\Lambda}_{\mathrm{func}}
\odot\boldsymbol{\delta}_{\mathrm{func}}
\right),
\label{eq:app_functional_forecast}
\end{equation}
where $\alpha_{\mathrm{func}}$ controls the residual correction
scale. The three outputs
$\mathbf{F}^{\mathrm{func}}$,
$\mathbf{F}^{\rightarrow}$, and
$\mathbf{F}^{\leftarrow}$ are passed to adaptive fusion.

\subsection{Context-Adaptive Forecast Fusion}
\label{app:fusion}

We detail the context construction, fusion weights, and output
recovery in Section~\ref{sec:fusion}. 

\paragraph{Context construction and fusion.}
For node $i$ and target patch $q$, the fusion context is
\begin{equation}
\mathbf{c}_{i,q}
=
\left[
\mathbf{e}_{i}
\Vert\mathbf{u}^{\mathrm{tod}}_{q}
\Vert\mathbf{u}^{\mathrm{dow}}_{q}
\Vert\mathbf{p}_{q}
\right]
\in\mathbb{R}^{4D},
\label{eq:app_fusion_context}
\end{equation}
where $\mathbf{e}_{i}\in\mathbb{R}^{D}$ is the shared node
embedding, and $\mathbf{u}^{\mathrm{tod}}_{q}$ and
$\mathbf{u}^{\mathrm{dow}}_{q}$ are fusion-specific calendar
embeddings at the target-patch start time. Their lookup tables
contain 288 time-of-day bins and seven day-of-week categories,
respectively. The positional encoding $\mathbf{p}_{q}\in\mathbb{R}^{D}$
uses the zero-based target-patch position $q-1$:
\begin{equation}
[\mathbf{p}_{q}]_{2m}
=
\sin\!\left(\frac{q-1}{10000^{2m/D}}\right),
\qquad
[\mathbf{p}_{q}]_{2m+1}
=
\cos\!\left(\frac{q-1}{10000^{2m/D}}\right),
\label{eq:app_fusion_position}
\end{equation}
for nonnegative integers $m$ whose feature indices are below $D$.
Calendar embeddings describe the target time, whereas
$\mathbf{p}_{q}$ identifies its position within the forecast.

The fusion MLP contains two affine layers with GELU activation
and dropout between them. Its three outputs are normalized
by softmax and applied to the branch representations as
specified in Equation~\ref{eq:context_fusion}.
Each weight is a scalar for one node, target patch, and branch,
broadcast across $D_{\mathrm{f}}$ features.

\paragraph{Prediction and overlap recovery.}
A shared two-layer prediction head maps each fused representation
to a length-$\ell$ output patch:
\begin{equation}
\hat{\mathbf{y}}^{\mathrm{patch}}_{i,q}
=
\mathbf{W}_{2}\,
\operatorname{ReLU}\!\left(
\mathbf{W}_{1}
\operatorname{ReLU}(\mathbf{F}^{\mathrm{fuse}}_{i,q})
+\mathbf{b}_{1}
\right)
+\mathbf{b}_{2},
\label{eq:app_fusion_prediction}
\end{equation}
where $\hat{\mathbf{y}}^{\mathrm{patch}}_{i,q}\in\mathbb{R}^{\ell}$,
$\mathbf{W}_{1}\in\mathbb{R}^{D_o\times D_{\mathrm{f}}}$,
$\mathbf{W}_{2}\in\mathbb{R}^{\ell\times D_o}$,
and $D_o$ is the prediction-head hidden dimension.
The vectors $\mathbf{b}_{1}$ and $\mathbf{b}_{2}$ are biases. Target patches continue the historical patch grid.
Using zero-based positions relative to the input-window start,
the start of target patch $q$ is $\kappa_q=(P+q-1)s$,
with patches included while $\kappa_q<L+H$.
For forecasting step $h\in\{1,\ldots,H\}$, define
\[
\mathcal{I}_{h}
=
\left\{
q:0\leq L+h-1-\kappa_q<\ell
\right\}
\]
as the set of target patches covering that step.
The final forecast averages their predictions:
\begin{equation}
[\hat{\mathbf{Y}}_{t+1:t+H}]_{h,i}
=
\frac{
\displaystyle\sum_{q\in\mathcal{I}_{h}}
[\hat{\mathbf{y}}^{\mathrm{patch}}_{i,q}]_{L+h-\kappa_q}
}{
\max(1,|\mathcal{I}_{h}|)
}.
\label{eq:app_overlap_recovery}
\end{equation}
Only positions within the forecasting interval are retained,
yielding $\hat{\mathbf{Y}}_{t+1:t+H}\in\mathbb{R}^{H\times N}$.

\subsection{Learning Objectives}
\label{app:training_objectives}

\paragraph{Reconstruction and role consistency.}
Equation~\ref{eq:evolution_reconstruction} uses component
outputs before layer normalization. The smooth-$L_1$ loss
averages the element-wise penalty
$\phi(u)=u^2/2$ for $|u|<1$ and
$\phi(u)=|u|-1/2$ otherwise.
The shifted reconstruction is evaluated only where an
observed successor exists and is zero when $P=1$.
Gradients through the target $\mathbf{Z}$ are stopped;
the component outputs remain differentiable.

For direction $d$, let $\mathbf{L}_{i,p}^{d}$,
$\mathbf{E}_{i,p}^{d}$, and $\mathbf{o}_{i,p}^{d}$
denote the temporal state, propagation residual, and complete
graph contribution defined in Appendix~\ref{app:dual_graph}.
We combine the two directions as
$\mathbf{L}_{i,p}^{\Sigma}
=\mathbf{L}_{i,p}^{\rightarrow}
+\mathbf{L}_{i,p}^{\leftarrow}$,
with $\mathbf{E}^{\Sigma}$ and $\mathbf{o}^{\Sigma}$
defined analogously. For each component label $c\in\{B,A,R\}$, define
\begin{equation}
\operatorname{sim}_{c}(\mathbf{x},\mathbf{y})
=
\cos\!\left(
\mathcal{P}_{c}^{\mathrm{comp}}(\mathbf{x}),
\mathcal{P}_{c}^{\mathrm{phys}}(\mathbf{y})
\right),
\label{eq:app_projected_similarity}
\end{equation}
where $\mathcal{P}_{c}^{\mathrm{comp}}$ and
$\mathcal{P}_{c}^{\mathrm{phys}}$ are independent
$D$-dimensional linear maps initialized to the identity.
Role consistency is
\begin{equation}
\begin{aligned}
\mathcal{L}_{\mathrm{role}}
={}&-\rho_B\,\mathbb{E}_{i,p}
\operatorname{sim}_{B}
(\mathbf{B}_{i,p+1},\mathbf{L}_{i,p}^{\Sigma})\\
&-\rho_A\,\mathbb{E}_{i,p}
\operatorname{sim}_{A}
(\mathbf{A}_{i,p+1},\mathbf{o}_{i,p}^{\Sigma})\\
&-\rho_R\,\mathbb{E}_{i,p}
\operatorname{sim}_{R}
(\mathbf{R}_{i,p},\mathbf{E}_{i,p}^{\Sigma}),
\end{aligned}
\label{eq:app_role_consistency}
\end{equation}
where $\rho_B$, $\rho_A$, and $\rho_R$ are nonnegative
weights. The background and accumulation terms use
$p=\max(K_t-1,1),\ldots,P-1$; the release term uses
$p=1,\ldots,P$.

The first two terms associate next-patch background and
accumulation with the temporal and graph contributions
used to predict that patch. The third associates release
with the residual available for propagation at the current
patch. All comparisons are within the same node, with
gradients through both sides. These are learned semantic
constraints rather than assertions of physical conservation. Reconstruction and role consistency are activated by
$\chi(e)=\mathbb{I}[e\geq e_0]$, where $e_0$ is the activation
epoch. This indicator does not delay the other auxiliary terms.

\paragraph{Prediction supervision and auxiliary regularization.}
Let $\mathbf{Y}^{\star}$ denote the ground-truth future.
The final forecast is supervised by
$\mathcal{L}_{\mathrm{pred}}
=\operatorname{MAE}(\hat{\mathbf{Y}},\mathbf{Y}^{\star})$,
where MAE averages absolute errors over valid targets.
A shared auxiliary prediction head followed by overlap
recovery decodes each branch representation into
$\hat{\mathbf{Y}}^{b}$, giving
\begin{equation}
\mathcal{L}_{\mathrm{branch}}
=
\frac{1}{3}\sum_{b\in\mathcal{B}}
\operatorname{MAE}(\hat{\mathbf{Y}}^{b},\mathbf{Y}^{\star}).
\label{eq:app_branch_supervision}
\end{equation}
This supplies a direct prediction signal to each branch
before adaptive fusion. The mean historical next-state prediction of the two physical
paths is supervised against the refined history:
\begin{equation}
\mathcal{L}_{\mathrm{hist}}
=
\mathbb{E}_{i,p<P}
\frac{1}{D}
\left\|
\frac{\mathbf{L}_{i,p}^{\Sigma}
+\mathbf{o}_{i,p}^{\Sigma}}{2}
-\mathbf{H}^{(0)}_{i,p+1}
\right\|_2^2.
\label{eq:app_historical_consistency}
\end{equation}

To discourage redundant temporal and propagation
representations, we average $\mathbf{L}^{d}$ and
$\mathbf{E}^{d}$ over historical patches and subtract
each resulting vector's mean across features.
Denoting these centered vectors by
$\overline{\mathbf{L}}_{i}^{d}$ and
$\overline{\mathbf{E}}_{i}^{d}$, we use
\begin{equation}
\mathcal{L}_{\mathrm{sep}}
=
\frac{1}{2}
\sum_{d\in\{\rightarrow,\leftarrow\}}
\mathbb{E}_{i}
\left|
\cos\!\left(
\overline{\mathbf{L}}_{i}^{d},
\overline{\mathbf{E}}_{i}^{d}
\right)
\right|.
\label{eq:app_state_separation}
\end{equation}
This penalty acts within each physical direction rather
than directly separating the functional and physical branches.

Background smoothness penalizes second temporal differences.
Shock sparsity penalizes the mean activity of the scalar
gates $g^S_U$ and $g^S_V$ from Appendix~\ref{app:decomposition}:
\begin{equation}
\mathcal{L}_{\mathrm{bg}}
=
\mathbb{E}_{i,p\geq3}
\frac{
\|\mathbf{B}_{i,p}-2\mathbf{B}_{i,p-1}
+\mathbf{B}_{i,p-2}\|_1
}{D},
\qquad
\mathcal{L}_{\mathrm{shock}}
=
\mathbb{E}_{i,p}
\frac{g^S_{U,i,p}+g^S_{V,i,p}}{2}.
\label{eq:app_component_regularization}
\end{equation}
The shock penalty encourages selective gate activation,
without enforcing exact zeros in the normalized component. These terms form the auxiliary objective
\begin{equation}
\begin{aligned}
\mathcal{L}_{\mathrm{aux}}
={}&
\lambda_{\mathrm{branch}}\mathcal{L}_{\mathrm{branch}}
+\lambda_{\mathrm{hist}}\mathcal{L}_{\mathrm{hist}}
+\lambda_{\mathrm{sep}}\mathcal{L}_{\mathrm{sep}}\\
&+\lambda_{\mathrm{bg}}\mathcal{L}_{\mathrm{bg}}
+\lambda_{\mathrm{shock}}\mathcal{L}_{\mathrm{shock}},
\end{aligned}
\label{eq:app_auxiliary_objective}
\end{equation}
where all coefficients are nonnegative.
The complete training objective is given in
Equation~\ref{eq:overall_objective}.

\subsection{Tasks and Datasets}
\label{app:data}

\paragraph{Traffic forecasting.}
METR-LA and PEMS-BAY contain traffic-speed measurements collected every
five minutes from 207 and 325 sensors, respectively. Following the
standard protocol, we use the preceding 12 observations to predict the
next 12 steps and divide each dataset chronologically into
70\%/10\%/20\% training, validation, and test sets. Missing observations
are processed according to the protocol of each benchmark. All
normalization statistics are estimated exclusively from the training
set and then applied to the validation and test sets.

\paragraph{Water-quality forecasting.}
We collect and curate publicly available hourly dissolved-oxygen
observations from the USGS National Water Information System
(NWIS)\footnote{\url{https://waterdata.usgs.gov/}} to construct three
five-year datasets: BASIN030501 (Santee), BASIN031300 (Apalachicola),
and BASIN180102 (Klamath). Each dataset contains at least ten monitoring
stations, has no more than 10\% missing observations per station, and
includes at least $N-1$ directed river connections, ensuring consistent
temporal coverage and sufficient network connectivity.

\begin{table}[htbp]
\centering
\caption{Statistics of the water-quality datasets.}
\label{tab:water_dataset_statistics}

\vspace{4pt}

\scriptsize
\setlength{\tabcolsep}{13pt}
\renewcommand{\arraystretch}{0.90}

\begin{tabular}{@{}llccccc@{}}
\toprule
Dataset & Basin & Period & Sites & Edges
& Station-hours & Observed \\
\midrule
BASIN030501 & Santee
& 2014--2018 & 12 & 17 & $525{,}888$ & 98.66\% \\
BASIN031300 & Apalachicola
& 2017--2022 & 20 & 28 & $876{,}480$ & 96.44\% \\
BASIN180102 & Klamath
& 2020--2025 & 11 & 42 & $482{,}064$ & 98.15\% \\
\midrule
Total & -- & -- & 43 & 87
& $1{,}884{,}432$ & 97.50\% \\
\bottomrule
\end{tabular}
\end{table}

Invalid sentinels, nonnumeric records, infinite values, and negative
measurements are treated as missing. After quality control, 1,837,295
of the 1,884,432 station-hour entries remain observed. Missing entries
are filled independently within each station by forward filling followed
by backward filling, and the same processed series is used by all
methods. Each dataset contains 43,824 hourly timestamps and is divided
chronologically into 60\%/20\%/20\% training, validation, and test
periods. Directed physical graphs are constructed from downstream river-network
reachability. An edge $i\rightarrow j$ is retained when station $j$ is
downstream of station $i$ within 200 river kilometers; self-loops are
excluded. River-path distances define
the edge lengths and are converted to Gaussian weights when required. Dataset-level statistics are summarized in Table~\ref{tab:water_dataset_statistics}.
All models observe the preceding 48 hours and predict
$H\in\{48,96,144,192,240\}$ hours.

\subsection{Implementation Details}
\label{app:implementation}

\paragraph{Architecture and optimization.}
We use the same backbone across all five datasets, with the shared
hyperparameters summarized in Table~\ref{tab:hp-shared}. The physical
branch uses a two-patch temporal window and two spatial propagation
orders. The functional branch retains the top 16 relations per node,
with temporal-context conditioning enabled; shock-conditioned
modulation is used in the physical branch. All components are jointly
trained from scratch, without exogenous forecasting covariates.

\begin{table}[htbp]
\centering
\caption{Hyperparameters shared across all five datasets.}
\label{tab:hp-shared}

\vspace{4pt}

\scriptsize
\setlength{\tabcolsep}{10pt}
\renewcommand{\arraystretch}{0.90}

\begin{tabular}{@{}llll@{}}
\toprule
\multicolumn{2}{c}{Architecture}
& \multicolumn{2}{c}{Optimization} \\
\cmidrule(r){1-2}\cmidrule(l){3-4}
Hidden dimension & 32
& Optimizer & Adam \\
Forecast dimension & 256
& Learning rate & $2\times10^{-3}$ \\
Output hidden units & 512
& Weight decay & $10^{-5}$ \\
Attention heads & 4
& Adam $\epsilon$ & $10^{-8}$ \\
Dropout & 0.1
& Gradient clipping & 5.0 \\
Temporal span $K_t$ & 2
& LR schedule & MultiStep, $\gamma=0.5$ \\
Spatial orders $K_s$ & 2
& Early-stopping patience & 50 epochs \\
Relations per node & 16
& Checkpoint criterion & validation MAE \\
Fusion hidden units & 128
& Normalization & global $z$-score \\
Initial fusion weights & $(0.5,0.25,0.25)$
& Initialization & Xavier \\
\cmidrule(r){1-2}\cmidrule(l){3-4}
\multicolumn{4}{@{}l}{Auxiliary coefficients} \\
Branch $\lambda_{\mathrm{branch}}$ & 0.2
& Background $\lambda_{\mathrm{bg}}$ & $10^{-4}$ \\
History $\lambda_{\mathrm{hist}}$ & 0.2
& Shock $\lambda_{\mathrm{shock}}$ & $10^{-4}$ \\
Separation $\lambda_{\mathrm{sep}}$ & 0.2
& Shock threshold $\tau$ & 0.1 \\
\bottomrule
\end{tabular}
\end{table}

\paragraph{Dataset-specific settings.}
Table~\ref{tab:hp-dataset} reports the settings that vary across
datasets. Traffic forecasting uses 12 observed and 12 target steps,
whereas water-quality forecasting uses 48 observed hours and horizons
up to 240 hours. Patch geometry is adjusted to these temporal scales while the remaining architecture is unchanged. 
For water-quality forecasting, the patch length, stride, and curriculum
increment increase with the forecast horizon, as shown in
Table~\ref{tab:hp-horizon}. At 240 hours, the patch length remains 16
rather than increasing to 20, which preserves sufficient temporal
resolution for the 48-hour input. All other settings are shared across
horizons and basins.

\begin{table}[htbp]
\centering
\caption{Dataset-specific hyperparameters. Water-quality settings are
shown for the 192-hour horizon; the remaining horizons are given in
Table~\ref{tab:hp-horizon}.}
\label{tab:hp-dataset}

\vspace{4pt}

\scriptsize
\setlength{\tabcolsep}{6pt}
\renewcommand{\arraystretch}{0.90}

\begin{tabular}{@{}lccccc@{}}
\toprule
& METR-LA & PEMS-BAY
& BASIN030501 & BASIN031300 & BASIN180102 \\
\midrule
Nodes & 207 & 325 & 12 & 20 & 11 \\
History length & 12 & 12 & 48 & 48 & 48 \\
Forecast length & 12 & 12 & 192 & 192 & 192 \\
Patch length / stride & 4/2 & 3/3 & 16/8 & 16/8 & 16/8 \\
Relation channels & 4 & 2 & 4 & 4 & 4 \\
Batch size & 64 & 64 & 128 & 128 & 128 \\
Epoch budget & 80 & 50 & 40 & 40 & 40 \\
Warm-up epochs & 0 & 20 & 0 & 0 & 0 \\
Curriculum start / increment & 1 & 1 & 16 & 16 & 16 \\
Curriculum period (epochs) & 6 & 2 & 3 & 3 & 3 \\
LR milestones
& $12,24,\ldots,72$
& $26,32,\ldots,50$
& $6,12,\ldots,36$
& $6,12,\ldots,36$
& $6,12,\ldots,36$ \\
Activation epoch $e_0$ & 36 & 10 & 0 & 0 & 0 \\
Reconstruction $\lambda_{\mathrm{rec}}$
& 0.1 & 0.1 & 0.02 & 0.02 & 0.02 \\
Role consistency ($B/A/R$)
& 0.2 & 0.2 & 0.04 & 0.04 & 0.04 \\
Loss space
& original & original & normalized & normalized & normalized \\
Loss masking
& applied & applied & none & none & none \\
Parameters
& $898{,}566$ & $904{,}373$
& $876{,}750$ & $878{,}030$ & $876{,}590$ \\
\bottomrule
\end{tabular}
\end{table}

\begin{table}[htbp]
\centering
\caption{Water-quality settings by forecast horizon, shared across the
three basins.}
\label{tab:hp-horizon}

\vspace{4pt}

\scriptsize
\setlength{\tabcolsep}{16pt}
\renewcommand{\arraystretch}{0.90}

\begin{tabular}{@{}lccccc@{}}
\toprule
Forecast horizon (hours) & 48 & 96 & 144 & 192 & 240 \\
\midrule
Patch length / stride & 4/2 & 8/4 & 12/6 & 16/8 & 16/8 \\
Curriculum start / increment & 4 & 8 & 12 & 16 & 20 \\
\bottomrule
\end{tabular}
\end{table}

\paragraph{Training and evaluation.}
The forecast curriculum begins with the prefix specified in
Table~\ref{tab:hp-dataset} and extends it by the same number of steps at
each curriculum period until the complete horizon is supervised.
PEMS-BAY additionally uses a 20-epoch full-horizon warm-up before the
curriculum begins. Reconstruction and role-consistency losses are
activated at epoch $e_0$; all remaining terms are active throughout training. Experiments are conducted on a single NVIDIA RTX~5090 GPU. 


\paragraph{Parameter sensitivity.}
We examine three structural hyperparameters governing relation learning:
the temporal propagation span $K_t$, maximum spatial order $K_s$, and
functional Top-$K$ value $K$. Using a fix random seed, we vary one parameter at a
time on METR-LA and BASIN180102 at the 192-hour horizon while keeping
all other settings fixed. The shared default is
$(K_t,K_s,K)=(2,2,16)$.

\begin{table*}[htbp]
\centering
\caption{One-at-a-time parameter sensitivity on representative traffic
and water-quality settings. Gray rows denote the shared
default configuration.}
\label{tab:param_sensitivity}
\vspace{2pt}
\scriptsize
\setlength{\tabcolsep}{9pt}
\renewcommand{\arraystretch}{1.15}
\begin{tabular}{llcccc}
\toprule
\textbf{Parameter} &
\textbf{Value} &
\multicolumn{2}{c}{\textbf{METR-LA}} &
\multicolumn{2}{c}{\textbf{BASIN180102 ($H=192$)}} \\
\cmidrule(lr){3-4}
\cmidrule(lr){5-6}
& & MAE & RMSE & MAE & MSE \\
\midrule
$K_t$ & 1
& 2.8549 & 5.8625 & 0.7115 & 1.5168 \\
\rowcolor{gray!8}
$K_t$ & 2
& 2.8439 & 5.8280 & 0.7151 & 1.5289 \\
$K_t$ & 3
& 2.8575 & 5.8632 & 0.7105 & 1.5304 \\
$K_t$ & 4
& 2.8612 & 5.8803 & 0.7259 & 1.5625 \\
\addlinespace[2pt]
$K_s$ & 1
& 2.8725 & 5.9097 & 0.7173 & 1.5485 \\
\rowcolor{gray!8}
$K_s$ & 2
& 2.8439 & 5.8280 & 0.7151 & 1.5289 \\
$K_s$ & 3
& 2.8565 & 5.8787 & 0.7199 & 1.5553 \\
\addlinespace[2pt]
$K$ & 4
& 2.8584 & 5.8771 & 0.7116 & 1.5299 \\
$K$ & 8
& 2.8530 & 5.8705 & 0.7090 & 1.5191 \\
\rowcolor{gray!8}
$K$ & 16
& 2.8439 & 5.8280 & 0.7151 & 1.5289 \\
$K$ & 32
& 2.8492 & 5.8572 & 0.7139 & 1.5317 \\
\bottomrule
\end{tabular}
\end{table*}

Table~\ref{tab:param_sensitivity} shows limited sensitivity across the
tested ranges. On METR-LA, the shared default achieves the lowest MAE
and RMSE in all three sweeps, and the largest MAE increase is only
$1.01\%$. On BASIN180102, MAE changes range from $-0.85\%$ to
$+1.51\%$, while MSE changes remain between $-0.79\%$ and $+2.20\%$.
A long temporal span ($K_t=4$) produces the clearest degradation,
whereas $K_s=2$ performs consistently well on both systems. Performance
also varies modestly over $K\in\{4,8,16,32\}$, indicating that
functional relation selection does not rely on a narrowly tuned
sparsity level. Overall, the shared default provides a stable
cross-system configuration without dataset-specific tuning.


\paragraph{Computational complexity.}
We further compare CANDOR with all 16 baselines in terms of model size
and inference cost. For runtime profiling, all methods are evaluated in
FP32 on the same NVIDIA RTX~5090 using a common batch size of 64.
Each model is first warmed up for 20 forward passes, after which
inference latency is averaged over 100 forward passes. Trainable
parameter counts are obtained from the corresponding model
configurations.

\begin{figure}[htbp]
    \centering
    \includegraphics[width=0.98\textwidth]{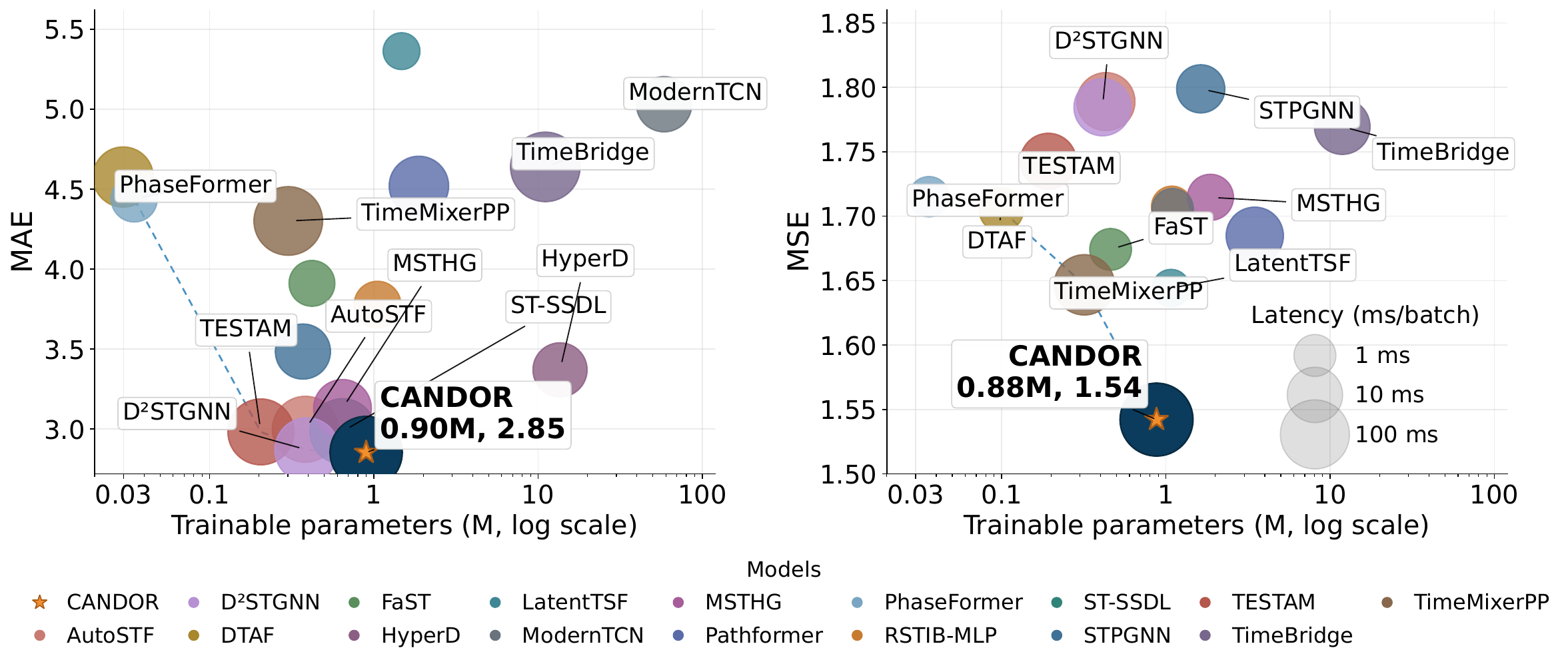}
    \caption{Accuracy-complexity trade-off on representative traffic and
water-quality datasets. The horizontal axis shows trainable parameters
on a logarithmic scale, the vertical axis reports forecasting error,
and bubble sizes encode per-batch inference latency using a
logarithmic mapping. Dashed lines mark
the parameter-accuracy Pareto frontier. METR-LA is shown on the left and BASIN180102 on the right. Points outside the displayed error range are omitted.} 
    \label{fig:complexity_tradeoff}
\end{figure}

Figure~\ref{fig:complexity_tradeoff} shows METR-LA and BASIN180102 at
the 192-hour horizon as representative traffic and water-quality
settings. On METR-LA, CANDOR achieves the lowest MAE of $2.854$ with
only $0.90$M trainable parameters. On BASIN180102, it achieves the
lowest MSE of $1.542$ with $0.88$M parameters. In both cases, CANDOR
lies at the high-accuracy end of the parameter-accuracy Pareto
frontier.

\subsection{Compared Methods and Evaluation Protocol}
\label{methods}

\paragraph{Compared methods.}
{\color{black}We compare CANDOR with 16 recent forecasting methods from two
complementary families. Spatiotemporal models explicitly exploit
cross-node interactions, whereas general time-series models provide
strong alternatives when spatial propagation is weak or local temporal
regularities dominate.}

\textbf{Spatiotemporal forecasting models.}
TESTAM combines temporal, static-graph, and dynamic-graph experts through
context-dependent routing \citep{lee2024testam}.
AutoSTF searches over decoupled spatial and temporal operators and uses
multi-patch transfer to capture multi-scale dependencies
\citep{lyu2025autostf}.
ST-SSDL models deviations from historical patterns through
self-supervised prototype learning and contrastive regularization
\citep{gao2025stssdl}.
{\color{black}D$^2$STGNN separates diffusion and inherent dynamics,
modeling propagation through dynamic graph convolution and local
evolution through recurrent and attention-based temporal networks
\citep{ShaoZWWXCJ22}.}
HyperD separates periodic structure from irregular residual dynamics
using frequency-aware representations \citep{shao2026hyperd}.
RSTIB-MLP distills a graph-based teacher into an efficient MLP under a
robust spatiotemporal information bottleneck \citep{rstib2025}.
STPGNN identifies pivotal nodes and models their distinct network
influence through pivotal graph convolutions \citep{kong2024stpgnn}.
MSTHG constructs multi-source spatial and temporal hypergraphs to
capture dynamic higher-order dependencies \citep{msthg2026}.
FaST combines temporal attention with state-space spatial modeling for
efficient long-range dependency learning \citep{fast2026}. These methods cover adaptive graphs, architecture search,
self-supervised learning, frequency modeling, distillation, pivotal-node
reasoning, hypergraphs, and state-space models.

\textbf{General time-series forecasting models.}
DTAF addresses temporal and spectral non-stationarity through
dual temporal-stabilization and frequency-differencing branches
\citep{dtaf2026}.
LatentTSF learns a structured latent representation and performs
forecasting within the resulting latent space \citep{latenttsf2026}.
ModernTCN enlarges the effective receptive field of temporal
convolutions using large-kernel designs \citep{moderntcn2024}.
Pathformer adaptively routes inputs through multi-scale temporal
pathways \citep{pathformer2024}.
PhaseFormer represents temporal patterns through compact phase-wise
tokens and lightweight cross-phase routing \citep{phaseformer2026}.
TimeBridge models short-term non-stationarity and long-term
cross-variable dependencies through integrated and cointegrated
attention \citep{timebridge2025}.
TimeMixer++ captures temporal patterns across multiple scales and
frequency resolutions \citep{wang2025timemixer++}.
These methods span convolutional, MLP-based, latent-space,
frequency-domain, and Transformer-style forecasting paradigms.

\paragraph{Evaluation protocol.}
All methods mentioned before use identical forecasting targets, input lengths, prediction
horizons, chronological splits, and evaluation pipelines, without
additional exogenous covariates. Each
experiment is repeated using the same three seeds, and the checkpoint with
the lowest validation MAE is retained for each run. For traffic forecasting, MAE is the primary metric, with RMSE and MAPE
reported as complementary measures. For water-quality forecasting, MSE
is the primary metric and MAE is complementary. All metrics are computed
after inverse transformation to the original scale.

\subsection{Additional Experiments}
\label{app:res}
{\color{black}Tables~\ref{tab:app_metrla}-\ref{tab:app_basin180102} provide complete
mean$\pm$std results for all 16 baselines and CANDOR; the main
paper reports the predefined top-eight subsets. Rankings use unrounded
means. These benchmark comparisons use the same forecasting information
and do not introduce additional exogenous covariates. CANDOR achieves
the lowest mean Overall MAE on both traffic datasets and the lowest
five-horizon mean MSE on all three water-quality datasets.
On METR-LA, even its worst-seed Overall MAE is lower than the best-seed
result of all 16 baselines. On PEMS-BAY, this separation holds for 15 of
the 16 baselines. For water quality, strict worst-to-best separation in
MSE holds for 235 of 240 basin-horizon-baseline comparisons, with the
five exceptions confined to BASIN030501 and BASIN031300.
These results indicate consistent improvements across seeds, datasets,
and forecast horizons.}

\begin{table}[htbp]
\centering
\caption{Complete METR-LA results for all 16 baselines and CANDOR, reported as
mean$\pm$std over three seeds. Best and second-best means are shown in
\textbf{bold} and \underline{underline}, respectively.}
\label{tab:app_metrla}
\vspace{2pt}
\scriptsize
\setlength{\tabcolsep}{3.2pt}
\renewcommand{\arraystretch}{0.85}

{\color{black}

\begin{tabular}{@{}lrrrrrrrrrrrr@{}}
\toprule
 & \multicolumn{3}{c}{\textbf{Overall}} & \multicolumn{3}{c}{\textbf{15 min}} & \multicolumn{3}{c}{\textbf{30 min}} & \multicolumn{3}{c}{\textbf{60 min}} \\
\cmidrule(lr){2-4} \cmidrule(lr){5-7} \cmidrule(lr){8-10} \cmidrule(lr){11-13}
\textbf{Method} & MAE & RMSE & MAPE & MAE & RMSE & MAPE & MAE & RMSE & MAPE & MAE & RMSE & MAPE \\
\midrule
\multirow{2}{*}{TESTAM} & \revised{2.982} & \revised{6.136} & 8.14\% & 2.670 & 5.206 & 6.89\% & \revised{3.017} & \revised{6.189} & 8.25\% & \revised{3.420} & \revised{7.215} & 9.86\% \\
 & $\pm$0.057 & $\pm$0.155 & $\pm$0.21 & $\pm$0.043 & $\pm$0.132 & $\pm$0.18 & $\pm$0.060 & $\pm$0.169 & $\pm$0.24 & $\pm$0.074 & $\pm$0.179 & $\pm$0.24 \\[1.6pt]
\multirow{2}{*}{AutoSTF} & 2.999 & 6.195 & 8.16\% & 2.692 & 5.231 & 6.97\% & 3.040 & 6.254 & 8.29\% & 3.427 & 7.292 & \revised{9.81\%} \\
 & $\pm$0.003 & $\pm$0.002 & $\pm$0.12 & $\pm$0.010 & $\pm$0.022 & $\pm$0.10 & $\pm$0.008 & $\pm$0.013 & $\pm$0.12 & $\pm$0.006 & $\pm$0.019 & $\pm$0.18 \\[1.6pt]
\multirow{2}{*}{ST-SSDL} & 2.986 & 6.191 & \revised{8.13\%} & \revised{2.643} & \revised{5.133} & \revised{6.80\%} & 3.021 & 6.225 & \revised{8.23\%} & 3.474 & 7.429 & 10.00\% \\
 & $\pm$0.115 & $\pm$0.298 & $\pm$0.28 & $\pm$0.077 & $\pm$0.208 & $\pm$0.23 & $\pm$0.107 & $\pm$0.279 & $\pm$0.27 & $\pm$0.183 & $\pm$0.427 & $\pm$0.40 \\[1.6pt]
\multirow{2}{*}{\revised{D$^2$STGNN}} & \revised{\underline{2.870}} & \revised{\underline{5.879}} & \revised{\underline{7.80\%}} & \revised{\underline{2.554}} & \revised{\underline{4.898}} & \revised{\underline{6.52\%}} & \revised{\underline{2.902}} & \revised{\underline{5.916}} & \revised{\underline{7.92\%}} & \revised{\underline{3.343}} & \revised{\underline{7.033}} & \revised{\underline{9.64\%}} \\
 & \revised{$\pm$0.002} & \revised{$\pm$0.010} & \revised{$\pm$0.02} & \revised{$\pm$0.002} & \revised{$\pm$0.004} & \revised{$\pm$0.01} & \revised{$\pm$0.001} & \revised{$\pm$0.012} & \revised{$\pm$0.01} & \revised{$\pm$0.012} & \revised{$\pm$0.032} & \revised{$\pm$0.07} \\[1.6pt]
\multirow{2}{*}{HyperD} & 3.370 & 6.721 & 10.29\% & 3.091 & 5.956 & 9.20\% & 3.451 & 6.832 & 10.57\% & 3.778 & 7.747 & 12.04\% \\
 & $\pm$0.006 & $\pm$0.019 & $\pm$0.03 & $\pm$0.004 & $\pm$0.011 & $\pm$0.02 & $\pm$0.007 & $\pm$0.018 & $\pm$0.05 & $\pm$0.007 & $\pm$0.030 & $\pm$0.04 \\[1.6pt]
\multirow{2}{*}{RSTIB-MLP} & 3.778 & 9.082 & 10.12\% & 3.221 & 7.434 & 8.36\% & 3.856 & 9.265 & 10.39\% & 4.522 & 10.836 & 12.41\% \\
 & $\pm$0.026 & $\pm$0.077 & $\pm$0.05 & $\pm$0.008 & $\pm$0.039 & $\pm$0.04 & $\pm$0.023 & $\pm$0.063 & $\pm$0.05 & $\pm$0.051 & $\pm$0.128 & $\pm$0.10 \\[1.6pt]
\multirow{2}{*}{STPGNN} & 3.485 & 6.888 & 10.12\% & 2.977 & 5.706 & 8.12\% & 3.498 & 6.911 & 10.18\% & 4.257 & 8.339 & 13.08\% \\
 & $\pm$0.068 & $\pm$0.128 & $\pm$0.38 & $\pm$0.045 & $\pm$0.112 & $\pm$0.31 & $\pm$0.073 & $\pm$0.136 & $\pm$0.43 & $\pm$0.104 & $\pm$0.173 & $\pm$0.52 \\[1.6pt]
\multirow{2}{*}{MSTHG} & 3.129 & 6.357 & 8.84\% & 2.790 & 5.388 & 7.42\% & 3.177 & 6.445 & 9.03\% & 3.602 & 7.460 & 10.71\% \\
 & $\pm$0.006 & $\pm$0.004 & $\pm$0.07 & $\pm$0.005 & $\pm$0.003 & $\pm$0.10 & $\pm$0.004 & $\pm$0.016 & $\pm$0.08 & $\pm$0.014 & $\pm$0.020 & $\pm$0.16 \\[1.6pt]
\multirow{2}{*}{FaST} & 3.911 & 8.932 & 10.38\% & 3.191 & 6.977 & 8.26\% & 3.912 & 8.850 & 10.45\% & 4.976 & 11.231 & 13.40\% \\
 & $\pm$0.021 & $\pm$0.023 & $\pm$0.05 & $\pm$0.010 & $\pm$0.012 & $\pm$0.01 & $\pm$0.017 & $\pm$0.017 & $\pm$0.04 & $\pm$0.042 & $\pm$0.046 & $\pm$0.10 \\[1.6pt]
\midrule
\multirow{2}{*}{DTAF} & 4.576 & 10.152 & 11.74\% & 3.558 & 7.880 & 8.86\% & 4.544 & 10.054 & 11.62\% & 6.088 & 12.819 & 16.06\% \\
 & $\pm$0.029 & $\pm$0.125 & $\pm$0.04 & $\pm$0.002 & $\pm$0.048 & $\pm$0.02 & $\pm$0.030 & $\pm$0.155 & $\pm$0.05 & $\pm$0.059 & $\pm$0.141 & $\pm$0.09 \\[1.6pt]
\multirow{2}{*}{LatentTSF} & 5.363 & 9.720 & 13.57\% & 4.229 & 7.957 & 10.44\% & 5.352 & 9.778 & 13.54\% & 6.987 & 11.709 & 18.10\% \\
 & $\pm$0.136 & $\pm$0.057 & $\pm$0.22 & $\pm$0.085 & $\pm$0.036 & $\pm$0.18 & $\pm$0.136 & $\pm$0.043 & $\pm$0.24 & $\pm$0.213 & $\pm$0.081 & $\pm$0.31 \\[1.6pt]
\multirow{2}{*}{ModernTCN} & 5.031 & 10.548 & 12.68\% & 3.941 & 8.221 & 9.72\% & 5.005 & 10.449 & 12.57\% & 6.645 & 13.276 & 17.07\% \\
 & $\pm$0.010 & $\pm$0.009 & $\pm$0.01 & $\pm$0.023 & $\pm$0.031 & $\pm$0.01 & $\pm$0.031 & $\pm$0.041 & $\pm$0.01 & $\pm$0.022 & $\pm$0.029 & $\pm$0.01 \\[1.6pt]
\multirow{2}{*}{Pathformer} & 4.520 & 10.226 & 11.44\% & 3.529 & 7.889 & 8.74\% & 4.494 & 10.186 & 11.36\% & 5.952 & 12.798 & 15.39\% \\
 & $\pm$0.012 & $\pm$0.012 & $\pm$0.01 & $\pm$0.010 & $\pm$0.006 & $\pm$0.02 & $\pm$0.015 & $\pm$0.027 & $\pm$0.01 & $\pm$0.022 & $\pm$0.030 & $\pm$0.04 \\[1.6pt]
\multirow{2}{*}{PhaseFormer} & 4.439 & 9.562 & 11.82\% & 3.456 & 7.260 & 8.89\% & 4.371 & 9.386 & 11.59\% & 5.917 & 12.271 & 16.18\% \\
 & $\pm$0.010 & $\pm$0.009 & $\pm$0.03 & $\pm$0.010 & $\pm$0.012 & $\pm$0.03 & $\pm$0.004 & $\pm$0.007 & $\pm$0.05 & $\pm$0.016 & $\pm$0.004 & $\pm$0.03 \\[1.6pt]
\multirow{2}{*}{TimeBridge} & 4.639 & 10.025 & 12.38\% & 3.740 & 7.941 & 9.86\% & 4.582 & 9.864 & 12.24\% & 5.999 & 12.522 & 16.22\% \\
 & $\pm$0.747 & $\pm$1.137 & $\pm$2.22 & $\pm$0.750 & $\pm$1.367 & $\pm$2.29 & $\pm$0.734 & $\pm$1.135 & $\pm$2.16 & $\pm$0.725 & $\pm$0.959 & $\pm$2.06 \\[1.6pt]
\multirow{2}{*}{TimeMixer++} & 4.301 & 9.143 & 11.24\% & 3.403 & 7.050 & 8.71\% & 4.261 & 9.007 & 11.18\% & 5.659 & 11.632 & 15.02\% \\
 & $\pm$0.030 & $\pm$0.037 & $\pm$0.15 & $\pm$0.015 & $\pm$0.014 & $\pm$0.08 & $\pm$0.028 & $\pm$0.030 & $\pm$0.13 & $\pm$0.052 & $\pm$0.061 & $\pm$0.26 \\[1.6pt]
\midrule
\multirow{2}{*}{CANDOR (Ours)} & \textbf{2.854} & \textbf{5.860} & \textbf{7.77\%} & \textbf{2.548} & \textbf{4.892} & \textbf{6.47\%} & \textbf{2.889} & \textbf{5.891} & \textbf{7.86\%} & \textbf{3.300} & \textbf{6.991} & \textbf{9.62\%} \\
 & $\pm$0.009 & $\pm$0.028 & $\pm$0.03 & $\pm$0.008 & $\pm$0.024 & $\pm$0.06 & $\pm$0.011 & $\pm$0.029 & $\pm$0.05 & $\pm$0.004 & $\pm$0.041 & $\pm$0.05 \\[1.6pt]
\bottomrule
\end{tabular}
}
\end{table}

\begin{table}[htbp]
\centering
\caption{Complete PEMS-BAY results for all 16 baselines and CANDOR, reported as
mean$\pm$std over three seeds. Best and second-best means are shown in
\textbf{bold} and \underline{underline}, respectively.}
\label{tab:app_pemsbay}
\vspace{2pt}
\scriptsize
\setlength{\tabcolsep}{3.2pt}
\renewcommand{\arraystretch}{0.85}

{\color{black}

\begin{tabular}{@{}lrrrrrrrrrrrr@{}}
\toprule
 & \multicolumn{3}{c}{\textbf{Overall}} & \multicolumn{3}{c}{\textbf{15 min}} & \multicolumn{3}{c}{\textbf{30 min}} & \multicolumn{3}{c}{\textbf{60 min}} \\
\cmidrule(lr){2-4} \cmidrule(lr){5-7} \cmidrule(lr){8-10} \cmidrule(lr){11-13}
\textbf{Method} & MAE & RMSE & MAPE & MAE & RMSE & MAPE & MAE & RMSE & MAPE & MAE & RMSE & MAPE \\
\midrule
\multirow{2}{*}{TESTAM} & \revised{1.530} & \revised{3.566} & \revised{3.44\%} & \revised{1.297} & 2.781 & 2.74\% & \revised{1.593} & \revised{3.681} & \revised{3.59\%} & \underline{1.849} & \revised{4.320} & \underline{4.34\%} \\
 & $\pm$0.004 & $\pm$0.010 & $\pm$0.01 & $\pm$0.004 & $\pm$0.022 & $\pm$0.01 & $\pm$0.005 & $\pm$0.016 & $\pm$0.02 & $\pm$0.004 & $\pm$0.003 & $\pm$0.02 \\[1.6pt]
\multirow{2}{*}{AutoSTF} & 1.548 & 3.599 & 3.44\% & 1.298 & \revised{2.758} & 2.72\% & 1.613 & 3.712 & 3.61\% & 1.881 & 4.388 & 4.36\% \\
 & $\pm$0.010 & $\pm$0.019 & $\pm$0.02 & $\pm$0.004 & $\pm$0.011 & $\pm$0.02 & $\pm$0.011 & $\pm$0.020 & $\pm$0.03 & $\pm$0.017 & $\pm$0.026 & $\pm$0.03 \\[1.6pt]
\multirow{2}{*}{ST-SSDL} & 1.579 & 3.677 & 3.51\% & 1.304 & 2.765 & \revised{2.70\%} & 1.642 & 3.762 & 3.65\% & 1.946 & 4.555 & 4.55\% \\
 & $\pm$0.006 & $\pm$0.013 & $\pm$0.02 & $\pm$0.003 & $\pm$0.011 & $\pm$0.01 & $\pm$0.006 & $\pm$0.018 & $\pm$0.02 & $\pm$0.010 & $\pm$0.011 & $\pm$0.04 \\[1.6pt]
\multirow{2}{*}{\revised{D$^2$STGNN}} & \revised{\underline{1.514}} & \revised{\underline{3.500}} & \revised{\underline{3.41\%}} & \revised{\textbf{1.256}} & \revised{\textbf{2.636}} & \revised{\underline{2.63\%}} & \revised{\underline{1.569}} & \revised{\underline{3.582}} & \revised{\underline{3.55\%}} & \revised{1.861} & \revised{\underline{4.312}} & \revised{4.37\%} \\
 & \revised{$\pm$0.004} & \revised{$\pm$0.008} & \revised{$\pm$0.02} & \revised{$\pm$0.001} & \revised{$\pm$0.006} & \revised{$\pm$0.02} & \revised{$\pm$0.002} & \revised{$\pm$0.008} & \revised{$\pm$0.02} & \revised{$\pm$0.008} & \revised{$\pm$0.020} & \revised{$\pm$0.03} \\[1.6pt]
\multirow{2}{*}{HyperD} & 1.796 & 4.275 & 4.45\% & 1.643 & 3.889 & 4.07\% & 1.856 & 4.363 & 4.62\% & 2.109 & 4.948 & 5.24\% \\
 & $\pm$0.012 & $\pm$0.042 & $\pm$0.02 & $\pm$0.014 & $\pm$0.046 & $\pm$0.03 & $\pm$0.013 & $\pm$0.048 & $\pm$0.02 & $\pm$0.011 & $\pm$0.026 & $\pm$0.01 \\[1.6pt]
\multirow{2}{*}{RSTIB-MLP} & 1.600 & 3.716 & 3.61\% & 1.348 & 2.870 & 2.84\% & 1.665 & 3.832 & 3.77\% & 1.940 & 4.532 & 4.60\% \\
 & $\pm$0.001 & $\pm$0.015 & $\pm$0.01 & $\pm$0.003 & $\pm$0.006 & $\pm$0.00 & $\pm$0.001 & $\pm$0.016 & $\pm$0.01 & $\pm$0.004 & $\pm$0.010 & $\pm$0.04 \\[1.6pt]
\multirow{2}{*}{STPGNN} & 1.997 & 4.487 & 4.75\% & 1.457 & 3.079 & 3.04\% & 2.016 & 4.409 & 4.68\% & 2.754 & 5.964 & 7.14\% \\
 & $\pm$0.109 & $\pm$0.306 & $\pm$0.47 & $\pm$0.026 & $\pm$0.081 & $\pm$0.07 & $\pm$0.089 & $\pm$0.207 & $\pm$0.35 & $\pm$0.221 & $\pm$0.567 & $\pm$1.03 \\[1.6pt]
\multirow{2}{*}{MSTHG} & 1.586 & 3.683 & 3.59\% & 1.323 & 2.816 & 2.82\% & 1.651 & 3.801 & 3.75\% & 1.935 & 4.490 & 4.56\% \\
 & $\pm$0.005 & $\pm$0.020 & $\pm$0.00 & $\pm$0.004 & $\pm$0.017 & $\pm$0.01 & $\pm$0.005 & $\pm$0.025 & $\pm$0.01 & $\pm$0.005 & $\pm$0.021 & $\pm$0.00 \\[1.6pt]
\multirow{2}{*}{FaST} & 1.720 & 4.153 & 3.81\% & 1.366 & 2.983 & 2.86\% & 1.762 & 4.147 & 3.92\% & 2.232 & 5.368 & 5.13\% \\
 & $\pm$0.004 & $\pm$0.026 & $\pm$0.00 & $\pm$0.003 & $\pm$0.023 & $\pm$0.00 & $\pm$0.004 & $\pm$0.024 & $\pm$0.00 & $\pm$0.007 & $\pm$0.029 & $\pm$0.01 \\[1.6pt]
\midrule
\multirow{2}{*}{DTAF} & 2.083 & 5.055 & 4.59\% & 1.508 & 3.303 & 3.11\% & 2.077 & 4.875 & 4.54\% & 2.944 & 6.908 & 6.78\% \\
 & $\pm$0.007 & $\pm$0.011 & $\pm$0.05 & $\pm$0.010 & $\pm$0.005 & $\pm$0.04 & $\pm$0.007 & $\pm$0.011 & $\pm$0.04 & $\pm$0.004 & $\pm$0.016 & $\pm$0.08 \\[1.6pt]
\multirow{2}{*}{LatentTSF} & 2.544 & 5.034 & 5.93\% & 1.878 & 3.559 & 4.11\% & 2.532 & 4.938 & 5.89\% & 3.511 & 6.609 & 8.53\% \\
 & $\pm$0.021 & $\pm$0.009 & $\pm$0.04 & $\pm$0.008 & $\pm$0.015 & $\pm$0.04 & $\pm$0.019 & $\pm$0.009 & $\pm$0.04 & $\pm$0.045 & $\pm$0.009 & $\pm$0.04 \\[1.6pt]
\multirow{2}{*}{ModernTCN} & 2.152 & 5.088 & 4.76\% & 1.580 & 3.396 & 3.28\% & 2.153 & 4.918 & 4.73\% & 3.014 & 6.907 & 6.98\% \\
 & $\pm$0.001 & $\pm$0.001 & $\pm$0.01 & $\pm$0.001 & $\pm$0.001 & $\pm$0.01 & $\pm$0.001 & $\pm$0.002 & $\pm$0.01 & $\pm$0.001 & $\pm$0.002 & $\pm$0.01 \\[1.6pt]
\multirow{2}{*}{Pathformer} & 1.911 & 4.597 & 4.23\% & 1.460 & 3.220 & 3.07\% & 1.923 & 4.504 & 4.25\% & 2.596 & 6.108 & 5.95\% \\
 & $\pm$0.027 & $\pm$0.060 & $\pm$0.05 & $\pm$0.009 & $\pm$0.015 & $\pm$0.01 & $\pm$0.023 & $\pm$0.045 & $\pm$0.03 & $\pm$0.057 & $\pm$0.110 & $\pm$0.11 \\[1.6pt]
\multirow{2}{*}{PhaseFormer} & 2.118 & 4.981 & 4.79\% & 1.536 & 3.270 & 3.23\% & 2.112 & 4.807 & 4.72\% & 2.982 & 6.800 & 7.09\% \\
 & $\pm$0.012 & $\pm$0.022 & $\pm$0.06 & $\pm$0.015 & $\pm$0.028 & $\pm$0.08 & $\pm$0.017 & $\pm$0.026 & $\pm$0.10 & $\pm$0.009 & $\pm$0.023 & $\pm$0.05 \\[1.6pt]
\multirow{2}{*}{TimeBridge} & 1.911 & 4.581 & 4.26\% & 1.425 & 3.095 & 2.97\% & 1.921 & 4.475 & 4.26\% & 2.631 & 6.136 & 6.15\% \\
 & $\pm$0.002 & $\pm$0.001 & $\pm$0.00 & $\pm$0.001 & $\pm$0.001 & $\pm$0.00 & $\pm$0.002 & $\pm$0.001 & $\pm$0.01 & $\pm$0.003 & $\pm$0.002 & $\pm$0.01 \\[1.6pt]
\multirow{2}{*}{TimeMixer++} & 1.960 & 4.396 & 4.34\% & 1.478 & 3.073 & 3.09\% & 1.975 & 4.322 & 4.37\% & 2.674 & 5.832 & 6.14\% \\
 & $\pm$0.007 & $\pm$0.006 & $\pm$0.02 & $\pm$0.003 & $\pm$0.008 & $\pm$0.03 & $\pm$0.007 & $\pm$0.006 & $\pm$0.02 & $\pm$0.014 & $\pm$0.010 & $\pm$0.02 \\[1.6pt]
\midrule
\multirow{2}{*}{CANDOR (Ours)} & \textbf{1.508} & \textbf{3.500} & \textbf{3.34\%} & \revised{\underline{1.257}} & \revised{\underline{2.647}} & \textbf{2.59\%} & \textbf{1.565} & \textbf{3.582} & \textbf{3.46\%} & \textbf{1.846} & \textbf{4.308} & \textbf{4.31\%} \\
 & $\pm$0.008 & $\pm$0.029 & $\pm$0.04 & $\pm$0.003 & $\pm$0.008 & $\pm$0.02 & $\pm$0.009 & $\pm$0.025 & $\pm$0.04 & $\pm$0.014 & $\pm$0.049 & $\pm$0.07 \\[1.6pt]
\bottomrule
\end{tabular}
}

\end{table}

\begin{table}[htbp]
\centering
\caption{Complete BASIN030501 results for all 16 baselines and CANDOR, reported
as mean$\pm$std over three seeds. Best and second-best means are shown in \textbf{bold} and
\underline{underline}, respectively.}
\label{tab:app_basin030501}
\vspace{2pt}
\scriptsize
\setlength{\tabcolsep}{3.2pt}
\renewcommand{\arraystretch}{0.85}

{\color{black}

\begin{tabular}{@{}lrrrrrrrrrrrr@{}}
\toprule
 & \multicolumn{2}{c}{\textbf{48}} & \multicolumn{2}{c}{\textbf{96}} & \multicolumn{2}{c}{\textbf{144}} & \multicolumn{2}{c}{\textbf{192}} & \multicolumn{2}{c}{\textbf{240}} & \multicolumn{2}{c}{\textbf{Avg}} \\
\cmidrule(lr){2-3} \cmidrule(lr){4-5} \cmidrule(lr){6-7} \cmidrule(lr){8-9} \cmidrule(lr){10-11} \cmidrule(lr){12-13}
\textbf{Method} & MSE & MAE & MSE & MAE & MSE & MAE & MSE & MAE & MSE & MAE & MSE & MAE \\
\midrule
\multirow{2}{*}{TESTAM} & 0.422 & 0.428 & 0.489 & 0.464 & 0.612 & 0.529 & 0.691 & 0.573 & 0.748 & 0.601 & 0.592 & 0.519 \\
 & $\pm$0.015 & $\pm$0.009 & $\pm$0.008 & $\pm$0.013 & $\pm$0.058 & $\pm$0.031 & $\pm$0.033 & $\pm$0.022 & $\pm$0.041 & $\pm$0.023 & $\pm$0.028 & $\pm$0.019 \\[1.6pt]
\multirow{2}{*}{AutoSTF} & \revised{0.291} & 0.336 & 0.483 & 0.452 & 0.595 & 0.514 & 0.678 & 0.558 & 0.781 & 0.608 & 0.566 & 0.494 \\
 & $\pm$0.003 & $\pm$0.002 & $\pm$0.013 & $\pm$0.002 & $\pm$0.065 & $\pm$0.027 & $\pm$0.018 & $\pm$0.005 & $\pm$0.050 & $\pm$0.015 & $\pm$0.016 & $\pm$0.006 \\[1.6pt]
\multirow{2}{*}{ST-SSDL} & 0.339 & 0.377 & 0.548 & 0.495 & 0.700 & 0.574 & 0.853 & 0.639 & 0.973 & 0.695 & 0.683 & 0.556 \\
 & $\pm$0.013 & $\pm$0.006 & $\pm$0.027 & $\pm$0.011 & $\pm$0.038 & $\pm$0.014 & $\pm$0.015 & $\pm$0.014 & $\pm$0.131 & $\pm$0.050 & $\pm$0.026 & $\pm$0.005 \\[1.6pt]
\multirow{2}{*}{\revised{D$^2$STGNN}} & \revised{\textbf{0.285}} & \revised{\underline{0.331}} & \revised{0.476} & \revised{0.448} & \revised{0.650} & \revised{0.537} & \revised{0.739} & \revised{0.584} & \revised{0.850} & \revised{0.641} & \revised{0.600} & \revised{0.508} \\
 & \revised{$\pm$0.003} & \revised{$\pm$0.002} & \revised{$\pm$0.011} & \revised{$\pm$0.007} & \revised{$\pm$0.028} & \revised{$\pm$0.012} & \revised{$\pm$0.030} & \revised{$\pm$0.014} & \revised{$\pm$0.008} & \revised{$\pm$0.003} & \revised{$\pm$0.013} & \revised{$\pm$0.005} \\[1.6pt]
\multirow{2}{*}{HyperD} & 2.273 & 1.163 & 2.305 & 1.181 & 2.314 & 1.187 & 2.351 & 1.200 & 2.347 & 1.200 & 2.318 & 1.186 \\
 & $\pm$0.032 & $\pm$0.008 & $\pm$0.024 & $\pm$0.006 & $\pm$0.003 & $\pm$0.003 & $\pm$0.008 & $\pm$0.000 & $\pm$0.036 & $\pm$0.009 & $\pm$0.015 & $\pm$0.004 \\[1.6pt]
\multirow{2}{*}{RSTIB-MLP} & 0.370 & 0.409 & 0.500 & 0.477 & 0.608 & 0.531 & 0.685 & 0.569 & 0.752 & 0.602 & 0.583 & 0.518 \\
 & $\pm$0.004 & $\pm$0.004 & $\pm$0.004 & $\pm$0.002 & $\pm$0.005 & $\pm$0.003 & $\pm$0.002 & $\pm$0.001 & $\pm$0.001 & $\pm$0.000 & $\pm$0.002 & $\pm$0.001 \\[1.6pt]
\multirow{2}{*}{STPGNN} & 0.352 & 0.382 & 0.521 & 0.478 & 0.657 & 0.541 & 0.740 & 0.593 & 0.914 & 0.671 & 0.637 & 0.533 \\
 & $\pm$0.008 & $\pm$0.005 & $\pm$0.003 & $\pm$0.003 & $\pm$0.011 & $\pm$0.010 & $\pm$0.013 & $\pm$0.004 & $\pm$0.049 & $\pm$0.023 & $\pm$0.012 & $\pm$0.007 \\[1.6pt]
\multirow{2}{*}{MSTHG} & 0.347 & 0.359 & \underline{0.449} & 0.439 & \underline{0.558} & \underline{0.494} & \underline{0.629} & \underline{0.536} & \underline{0.706} & \underline{0.570} & \underline{0.538} & 0.480 \\
 & $\pm$0.016 & $\pm$0.003 & $\pm$0.006 & $\pm$0.002 & $\pm$0.007 & $\pm$0.004 & $\pm$0.017 & $\pm$0.007 & $\pm$0.017 & $\pm$0.005 & $\pm$0.009 & $\pm$0.003 \\[1.6pt]
\multirow{2}{*}{FaST} & 0.307 & 0.336 & 0.470 & \underline{0.429} & 0.604 & 0.495 & 0.711 & 0.547 & 0.802 & 0.589 & 0.579 & 0.479 \\
 & $\pm$0.002 & $\pm$0.000 & $\pm$0.003 & $\pm$0.001 & $\pm$0.001 & $\pm$0.001 & $\pm$0.002 & $\pm$0.001 & $\pm$0.001 & $\pm$0.000 & $\pm$0.001 & $\pm$0.000 \\[1.6pt]
\midrule
\multirow{2}{*}{DTAF} & 0.320 & 0.344 & 0.487 & 0.438 & 0.623 & 0.505 & 0.732 & 0.557 & 0.821 & 0.598 & 0.597 & 0.488 \\
 & $\pm$0.001 & $\pm$0.000 & $\pm$0.000 & $\pm$0.000 & $\pm$0.002 & $\pm$0.000 & $\pm$0.001 & $\pm$0.000 & $\pm$0.000 & $\pm$0.000 & $\pm$0.001 & $\pm$0.000 \\[1.6pt]
\multirow{2}{*}{LatentTSF} & 0.321 & 0.357 & 0.485 & 0.456 & 0.613 & 0.523 & 0.715 & 0.574 & 0.799 & 0.616 & 0.586 & 0.505 \\
 & $\pm$0.000 & $\pm$0.001 & $\pm$0.001 & $\pm$0.001 & $\pm$0.000 & $\pm$0.001 & $\pm$0.002 & $\pm$0.003 & $\pm$0.003 & $\pm$0.002 & $\pm$0.001 & $\pm$0.001 \\[1.6pt]
\multirow{2}{*}{ModernTCN} & 0.326 & 0.351 & 0.498 & 0.445 & 0.636 & 0.512 & 0.748 & 0.563 & 0.839 & 0.604 & 0.609 & 0.495 \\
 & $\pm$0.000 & $\pm$0.000 & $\pm$0.000 & $\pm$0.000 & $\pm$0.000 & $\pm$0.000 & $\pm$0.000 & $\pm$0.000 & $\pm$0.000 & $\pm$0.000 & $\pm$0.000 & $\pm$0.000 \\[1.6pt]
\multirow{2}{*}{Pathformer} & 0.311 & \revised{0.334} & 0.477 & 0.429 & 0.608 & 0.495 & 0.714 & 0.546 & 0.807 & 0.589 & 0.583 & \underline{0.479} \\
 & $\pm$0.001 & $\pm$0.001 & $\pm$0.001 & $\pm$0.000 & $\pm$0.002 & $\pm$0.000 & $\pm$0.002 & $\pm$0.000 & $\pm$0.003 & $\pm$0.001 & $\pm$0.001 & $\pm$0.000 \\[1.6pt]
\multirow{2}{*}{PhaseFormer} & 0.317 & 0.344 & 0.486 & 0.441 & 0.619 & 0.507 & 0.728 & 0.559 & 0.815 & 0.599 & 0.593 & 0.490 \\
 & $\pm$0.002 & $\pm$0.001 & $\pm$0.001 & $\pm$0.001 & $\pm$0.001 & $\pm$0.001 & $\pm$0.001 & $\pm$0.001 & $\pm$0.002 & $\pm$0.001 & $\pm$0.000 & $\pm$0.000 \\[1.6pt]
\multirow{2}{*}{TimeBridge} & 0.311 & 0.339 & 0.479 & 0.434 & 0.618 & 0.502 & 0.724 & 0.554 & 0.819 & 0.596 & 0.590 & 0.485 \\
 & $\pm$0.005 & $\pm$0.002 & $\pm$0.010 & $\pm$0.004 & $\pm$0.013 & $\pm$0.003 & $\pm$0.002 & $\pm$0.001 & $\pm$0.030 & $\pm$0.009 & $\pm$0.010 & $\pm$0.003 \\[1.6pt]
\multirow{2}{*}{TimeMixer++} & 0.313 & 0.342 & 0.477 & 0.436 & 0.611 & 0.502 & 0.719 & 0.554 & 0.811 & 0.598 & 0.586 & 0.486 \\
 & $\pm$0.000 & $\pm$0.001 & $\pm$0.001 & $\pm$0.001 & $\pm$0.008 & $\pm$0.003 & $\pm$0.010 & $\pm$0.003 & $\pm$0.002 & $\pm$0.001 & $\pm$0.000 & $\pm$0.000 \\[1.6pt]
\midrule
\multirow{2}{*}{CANDOR (Ours)} & \revised{\underline{0.289}} & \textbf{0.328} & \textbf{0.433} & \textbf{0.418} & \textbf{0.544} & \textbf{0.478} & \textbf{0.611} & \textbf{0.519} & \textbf{0.665} & \textbf{0.548} & \textbf{0.508} & \textbf{0.458} \\
 & $\pm$0.002 & $\pm$0.002 & $\pm$0.007 & $\pm$0.003 & $\pm$0.006 & $\pm$0.004 & $\pm$0.009 & $\pm$0.005 & $\pm$0.013 & $\pm$0.005 & $\pm$0.001 & $\pm$0.001 \\[1.6pt]
\bottomrule
\end{tabular}
}

\end{table}

\begin{table}[htbp]
\centering
\caption{Complete BASIN031300 results for all 16 baselines and CANDOR, reported
as mean$\pm$std over three seeds. Best and second-best means are shown in \textbf{bold} and
\underline{underline}, respectively.}
\label{tab:app_basin031300}
\vspace{2pt}
\scriptsize
\setlength{\tabcolsep}{3.2pt}
\renewcommand{\arraystretch}{0.85}

{\color{black}

\begin{tabular}{@{}lrrrrrrrrrrrr@{}}
\toprule
 & \multicolumn{2}{c}{\textbf{48}} & \multicolumn{2}{c}{\textbf{96}} & \multicolumn{2}{c}{\textbf{144}} & \multicolumn{2}{c}{\textbf{192}} & \multicolumn{2}{c}{\textbf{240}} & \multicolumn{2}{c}{\textbf{Avg}} \\
\cmidrule(lr){2-3} \cmidrule(lr){4-5} \cmidrule(lr){6-7} \cmidrule(lr){8-9} \cmidrule(lr){10-11} \cmidrule(lr){12-13}
\textbf{Method} & MSE & MAE & MSE & MAE & MSE & MAE & MSE & MAE & MSE & MAE & MSE & MAE \\
\midrule
\multirow{2}{*}{TESTAM} & 0.315 & 0.389 & 0.369 & 0.425 & 0.449 & 0.480 & 0.515 & 0.523 & 0.565 & 0.552 & 0.443 & 0.474 \\
 & $\pm$0.007 & $\pm$0.002 & $\pm$0.003 & $\pm$0.003 & $\pm$0.014 & $\pm$0.006 & $\pm$0.006 & $\pm$0.003 & $\pm$0.028 & $\pm$0.013 & $\pm$0.010 & $\pm$0.004 \\[1.6pt]
\multirow{2}{*}{AutoSTF} & 0.213 & 0.297 & 0.372 & 0.417 & 0.482 & 0.489 & 0.524 & 0.521 & 0.599 & 0.563 & 0.438 & 0.458 \\
 & $\pm$0.005 & $\pm$0.005 & $\pm$0.021 & $\pm$0.011 & $\pm$0.030 & $\pm$0.016 & $\pm$0.035 & $\pm$0.018 & $\pm$0.035 & $\pm$0.014 & $\pm$0.011 & $\pm$0.004 \\[1.6pt]
\multirow{2}{*}{ST-SSDL} & 0.244 & 0.329 & 0.446 & 0.462 & 0.550 & 0.528 & 0.652 & 0.584 & 0.775 & 0.644 & 0.533 & 0.509 \\
 & $\pm$0.010 & $\pm$0.008 & $\pm$0.052 & $\pm$0.028 & $\pm$0.033 & $\pm$0.016 & $\pm$0.063 & $\pm$0.025 & $\pm$0.070 & $\pm$0.032 & $\pm$0.013 & $\pm$0.008 \\[1.6pt]
\multirow{2}{*}{\revised{D$^2$STGNN}} & \revised{\textbf{0.203}} & \revised{\underline{0.289}} & \revised{0.368} & \revised{0.413} & \revised{0.507} & \revised{0.501} & \revised{0.575} & \revised{0.540} & \revised{0.704} & \revised{0.605} & \revised{0.471} & \revised{0.470} \\
 & \revised{$\pm$0.002} & \revised{$\pm$0.002} & \revised{$\pm$0.006} & \revised{$\pm$0.002} & \revised{$\pm$0.010} & \revised{$\pm$0.005} & \revised{$\pm$0.025} & \revised{$\pm$0.014} & \revised{$\pm$0.020} & \revised{$\pm$0.010} & \revised{$\pm$0.006} & \revised{$\pm$0.004} \\[1.6pt]
\multirow{2}{*}{HyperD} & 1.634 & 0.959 & 1.660 & 0.976 & 1.670 & 0.989 & 1.692 & 1.002 & 1.701 & 1.006 & 1.671 & 0.986 \\
 & $\pm$0.017 & $\pm$0.005 & $\pm$0.016 & $\pm$0.004 & $\pm$0.008 & $\pm$0.002 & $\pm$0.012 & $\pm$0.005 & $\pm$0.009 & $\pm$0.002 & $\pm$0.004 & $\pm$0.002 \\[1.6pt]
\multirow{2}{*}{RSTIB-MLP} & 0.314 & 0.397 & 0.430 & 0.470 & 0.502 & 0.511 & 0.557 & 0.544 & 0.600 & 0.568 & 0.480 & 0.498 \\
 & $\pm$0.008 & $\pm$0.007 & $\pm$0.002 & $\pm$0.002 & $\pm$0.005 & $\pm$0.004 & $\pm$0.002 & $\pm$0.002 & $\pm$0.003 & $\pm$0.001 & $\pm$0.003 & $\pm$0.003 \\[1.6pt]
\multirow{2}{*}{STPGNN} & 0.326 & 0.411 & 0.614 & 0.592 & 0.658 & 0.603 & 0.734 & 0.640 & 0.724 & 0.639 & 0.611 & 0.577 \\
 & $\pm$0.019 & $\pm$0.020 & $\pm$0.053 & $\pm$0.030 & $\pm$0.081 & $\pm$0.032 & $\pm$0.077 & $\pm$0.034 & $\pm$0.055 & $\pm$0.033 & $\pm$0.044 & $\pm$0.020 \\[1.6pt]
\multirow{2}{*}{MSTHG} & 0.227 & 0.310 & 0.371 & 0.417 & 0.469 & 0.481 & 0.529 & 0.519 & 0.585 & 0.552 & 0.436 & 0.456 \\
 & $\pm$0.004 & $\pm$0.001 & $\pm$0.005 & $\pm$0.004 & $\pm$0.013 & $\pm$0.006 & $\pm$0.004 & $\pm$0.002 & $\pm$0.008 & $\pm$0.006 & $\pm$0.004 & $\pm$0.002 \\[1.6pt]
\multirow{2}{*}{FaST} & \revised{0.213} & \revised{0.294} & \underline{0.350} & \underline{0.396} & \underline{0.441} & \underline{0.454} & \underline{0.507} & \underline{0.495} & \underline{0.557} & \underline{0.525} & \underline{0.414} & \underline{0.433} \\
 & $\pm$0.002 & $\pm$0.002 & $\pm$0.001 & $\pm$0.001 & $\pm$0.000 & $\pm$0.001 & $\pm$0.000 & $\pm$0.000 & $\pm$0.001 & $\pm$0.001 & $\pm$0.001 & $\pm$0.001 \\[1.6pt]
\midrule
\multirow{2}{*}{DTAF} & 0.237 & 0.314 & 0.377 & 0.414 & 0.471 & 0.472 & 0.539 & 0.512 & 0.591 & 0.542 & 0.443 & 0.451 \\
 & $\pm$0.001 & $\pm$0.001 & $\pm$0.001 & $\pm$0.001 & $\pm$0.000 & $\pm$0.000 & $\pm$0.000 & $\pm$0.000 & $\pm$0.001 & $\pm$0.000 & $\pm$0.000 & $\pm$0.000 \\[1.6pt]
\multirow{2}{*}{LatentTSF} & 0.248 & 0.332 & 0.381 & 0.428 & 0.468 & 0.482 & 0.530 & 0.520 & 0.578 & 0.547 & 0.441 & 0.462 \\
 & $\pm$0.000 & $\pm$0.000 & $\pm$0.000 & $\pm$0.000 & $\pm$0.000 & $\pm$0.000 & $\pm$0.000 & $\pm$0.000 & $\pm$0.000 & $\pm$0.000 & $\pm$0.000 & $\pm$0.000 \\[1.6pt]
\multirow{2}{*}{ModernTCN} & 0.253 & 0.330 & 0.392 & 0.426 & 0.486 & 0.483 & 0.553 & 0.521 & 0.606 & 0.551 & 0.458 & 0.462 \\
 & $\pm$0.000 & $\pm$0.000 & $\pm$0.000 & $\pm$0.000 & $\pm$0.000 & $\pm$0.000 & $\pm$0.000 & $\pm$0.000 & $\pm$0.000 & $\pm$0.000 & $\pm$0.000 & $\pm$0.000 \\[1.6pt]
\multirow{2}{*}{Pathformer} & 0.220 & 0.298 & 0.357 & 0.400 & 0.451 & 0.461 & 0.522 & 0.504 & 0.580 & 0.537 & 0.426 & 0.440 \\
 & $\pm$0.001 & $\pm$0.001 & $\pm$0.002 & $\pm$0.001 & $\pm$0.001 & $\pm$0.000 & $\pm$0.001 & $\pm$0.001 & $\pm$0.003 & $\pm$0.001 & $\pm$0.001 & $\pm$0.000 \\[1.6pt]
\multirow{2}{*}{PhaseFormer} & 0.231 & 0.312 & 0.369 & 0.411 & 0.465 & 0.470 & 0.532 & 0.510 & 0.584 & 0.539 & 0.436 & 0.448 \\
 & $\pm$0.004 & $\pm$0.003 & $\pm$0.001 & $\pm$0.000 & $\pm$0.002 & $\pm$0.002 & $\pm$0.003 & $\pm$0.002 & $\pm$0.003 & $\pm$0.002 & $\pm$0.001 & $\pm$0.001 \\[1.6pt]
\multirow{2}{*}{TimeBridge} & 0.222 & 0.301 & 0.365 & 0.406 & 0.468 & 0.468 & 0.527 & 0.505 & 0.593 & 0.539 & 0.435 & 0.444 \\
 & $\pm$0.001 & $\pm$0.001 & $\pm$0.006 & $\pm$0.004 & $\pm$0.012 & $\pm$0.005 & $\pm$0.001 & $\pm$0.001 & $\pm$0.001 & $\pm$0.001 & $\pm$0.003 & $\pm$0.002 \\[1.6pt]
\multirow{2}{*}{TimeMixer++} & 0.217 & 0.299 & 0.354 & 0.400 & 0.446 & 0.460 & 0.512 & 0.500 & 0.559 & 0.528 & 0.418 & 0.437 \\
 & $\pm$0.002 & $\pm$0.001 & $\pm$0.002 & $\pm$0.002 & $\pm$0.002 & $\pm$0.001 & $\pm$0.002 & $\pm$0.002 & $\pm$0.002 & $\pm$0.001 & $\pm$0.000 & $\pm$0.000 \\[1.6pt]
\midrule
\multirow{2}{*}{CANDOR (Ours)} & \revised{\underline{0.205}} & \textbf{0.287} & \textbf{0.331} & \textbf{0.388} & \textbf{0.420} & \textbf{0.449} & \textbf{0.487} & \textbf{0.493} & \textbf{0.528} & \textbf{0.520} & \textbf{0.394} & \textbf{0.428} \\
 & $\pm$0.004 & $\pm$0.003 & $\pm$0.004 & $\pm$0.003 & $\pm$0.006 & $\pm$0.005 & $\pm$0.006 & $\pm$0.004 & $\pm$0.008 & $\pm$0.006 & $\pm$0.005 & $\pm$0.004 \\[1.6pt]
\bottomrule
\end{tabular}
}
\end{table}

\begin{table}[htbp]
\centering
\caption{Complete BASIN180102 results for all 16 baselines and CANDOR, reported
as mean$\pm$std over three seeds. Best and second-best means are shown in \textbf{bold} and
\underline{underline}, respectively.}
\label{tab:app_basin180102}
\vspace{2pt}
\scriptsize
\setlength{\tabcolsep}{3.2pt}
\renewcommand{\arraystretch}{0.85}

{\color{black}

\begin{tabular}{@{}lrrrrrrrrrrrr@{}}
\toprule
 & \multicolumn{2}{c}{\textbf{48}} & \multicolumn{2}{c}{\textbf{96}} & \multicolumn{2}{c}{\textbf{144}} & \multicolumn{2}{c}{\textbf{192}} & \multicolumn{2}{c}{\textbf{240}} & \multicolumn{2}{c}{\textbf{Avg}} \\
\cmidrule(lr){2-3} \cmidrule(lr){4-5} \cmidrule(lr){6-7} \cmidrule(lr){8-9} \cmidrule(lr){10-11} \cmidrule(lr){12-13}
\textbf{Method} & MSE & MAE & MSE & MAE & MSE & MAE & MSE & MAE & MSE & MAE & MSE & MAE \\
\midrule
\multirow{2}{*}{TESTAM} & 0.900 & 0.551 & 1.215 & 0.622 & 1.598 & 0.733 & 1.743 & 0.788 & 2.066 & 0.865 & 1.505 & 0.712 \\
 & $\pm$0.005 & $\pm$0.002 & $\pm$0.104 & $\pm$0.015 & $\pm$0.148 & $\pm$0.024 & $\pm$0.099 & $\pm$0.003 & $\pm$0.244 & $\pm$0.011 & $\pm$0.108 & $\pm$0.008 \\[1.6pt]
\multirow{2}{*}{AutoSTF} & 0.750 & 0.455 & 1.142 & 0.592 & 1.489 & 0.703 & 1.789 & 0.783 & 2.208 & 0.877 & 1.476 & 0.682 \\
 & $\pm$0.030 & $\pm$0.008 & $\pm$0.086 & $\pm$0.024 & $\pm$0.094 & $\pm$0.018 & $\pm$0.195 & $\pm$0.043 & $\pm$0.129 & $\pm$0.021 & $\pm$0.032 & $\pm$0.008 \\[1.6pt]
\multirow{2}{*}{ST-SSDL} & 1.090 & 0.635 & 1.704 & 0.811 & 2.171 & 0.956 & 3.077 & 1.121 & 3.701 & 1.287 & 2.349 & 0.962 \\
 & $\pm$0.056 & $\pm$0.037 & $\pm$0.228 & $\pm$0.066 & $\pm$0.335 & $\pm$0.127 & $\pm$0.690 & $\pm$0.156 & $\pm$0.406 & $\pm$0.145 & $\pm$0.093 & $\pm$0.032 \\[1.6pt]
\multirow{2}{*}{\revised{D$^2$STGNN}} & \revised{0.752} & \revised{0.453} & \revised{1.185} & \revised{0.601} & \revised{1.648} & \revised{0.734} & \revised{1.784} & \revised{0.798} & \revised{2.292} & \revised{0.925} & \revised{1.532} & \revised{0.702} \\
 & \revised{$\pm$0.022} & \revised{$\pm$0.004} & \revised{$\pm$0.033} & \revised{$\pm$0.010} & \revised{$\pm$0.033} & \revised{$\pm$0.007} & \revised{$\pm$0.127} & \revised{$\pm$0.033} & \revised{$\pm$0.102} & \revised{$\pm$0.021} & \revised{$\pm$0.047} & \revised{$\pm$0.012} \\[1.6pt]
\multirow{2}{*}{HyperD} & 2.956 & 0.990 & 3.107 & 1.047 & 3.251 & 1.105 & 3.447 & 1.168 & 3.487 & 1.183 & 3.250 & 1.099 \\
 & $\pm$0.024 & $\pm$0.008 & $\pm$0.028 & $\pm$0.005 & $\pm$0.036 & $\pm$0.005 & $\pm$0.075 & $\pm$0.026 & $\pm$0.077 & $\pm$0.020 & $\pm$0.023 & $\pm$0.006 \\[1.6pt]
\multirow{2}{*}{RSTIB-MLP} & 0.921 & 0.606 & 1.214 & 0.672 & 1.486 & 0.743 & 1.707 & 0.802 & 1.882 & 0.848 & 1.442 & 0.734 \\
 & $\pm$0.020 & $\pm$0.013 & $\pm$0.019 & $\pm$0.012 & $\pm$0.030 & $\pm$0.017 & $\pm$0.001 & $\pm$0.001 & $\pm$0.014 & $\pm$0.007 & $\pm$0.007 & $\pm$0.005 \\[1.6pt]
\multirow{2}{*}{STPGNN} & 1.088 & 0.627 & 1.508 & 0.746 & 1.903 & 0.839 & 1.799 & 0.822 & 2.103 & 0.914 & 1.680 & 0.790 \\
 & $\pm$0.128 & $\pm$0.043 & $\pm$0.201 & $\pm$0.036 & $\pm$0.237 & $\pm$0.056 & $\pm$0.045 & $\pm$0.015 & $\pm$0.128 & $\pm$0.051 & $\pm$0.018 & $\pm$0.014 \\[1.6pt]
\multirow{2}{*}{MSTHG} & 0.748 & 0.460 & \underline{1.085} & 0.591 & 1.404 & 0.688 & 1.715 & 0.775 & 1.912 & 0.835 & 1.373 & 0.670 \\
 & $\pm$0.011 & $\pm$0.003 & $\pm$0.013 & $\pm$0.005 & $\pm$0.026 & $\pm$0.002 & $\pm$0.116 & $\pm$0.023 & $\pm$0.045 & $\pm$0.010 & $\pm$0.025 & $\pm$0.003 \\[1.6pt]
\multirow{2}{*}{FaST} & \underline{0.725} & 0.441 & 1.096 & 0.564 & \underline{1.390} & \underline{0.652} & 1.674 & 0.732 & 1.879 & \underline{0.789} & 1.353 & \underline{0.636} \\
 & $\pm$0.001 & $\pm$0.001 & $\pm$0.011 & $\pm$0.003 & $\pm$0.005 & $\pm$0.001 & $\pm$0.020 & $\pm$0.004 & $\pm$0.024 & $\pm$0.003 & $\pm$0.002 & $\pm$0.001 \\[1.6pt]
\midrule
\multirow{2}{*}{DTAF} & 0.743 & 0.451 & 1.111 & 0.578 & 1.429 & 0.672 & 1.705 & 0.748 & 1.924 & 0.810 & 1.382 & 0.652 \\
 & $\pm$0.002 & $\pm$0.001 & $\pm$0.005 & $\pm$0.001 & $\pm$0.000 & $\pm$0.000 & $\pm$0.004 & $\pm$0.000 & $\pm$0.002 & $\pm$0.000 & $\pm$0.001 & $\pm$0.000 \\[1.6pt]
\multirow{2}{*}{LatentTSF} & 0.739 & 0.473 & 1.091 & 0.606 & 1.394 & 0.701 & \underline{1.644} & 0.772 & \underline{1.846} & 0.835 & \underline{1.343} & 0.678 \\
 & $\pm$0.002 & $\pm$0.001 & $\pm$0.002 & $\pm$0.004 & $\pm$0.007 & $\pm$0.008 & $\pm$0.003 & $\pm$0.000 & $\pm$0.002 & $\pm$0.003 & $\pm$0.002 & $\pm$0.003 \\[1.6pt]
\multirow{2}{*}{ModernTCN} & 0.748 & 0.461 & 1.112 & 0.585 & 1.433 & 0.678 & 1.705 & 0.753 & 1.925 & 0.814 & 1.385 & 0.658 \\
 & $\pm$0.000 & $\pm$0.000 & $\pm$0.000 & $\pm$0.000 & $\pm$0.000 & $\pm$0.000 & $\pm$0.000 & $\pm$0.000 & $\pm$0.000 & $\pm$0.000 & $\pm$0.000 & $\pm$0.000 \\[1.6pt]
\multirow{2}{*}{Pathformer} & 0.732 & \underline{0.440} & 1.107 & \underline{0.563} & 1.408 & 0.653 & 1.685 & 0.731 & 1.890 & 0.791 & 1.364 & 0.636 \\
 & $\pm$0.003 & $\pm$0.001 & $\pm$0.006 & $\pm$0.002 & $\pm$0.010 & $\pm$0.002 & $\pm$0.014 & $\pm$0.001 & $\pm$0.009 & $\pm$0.001 & $\pm$0.004 & $\pm$0.000 \\[1.6pt]
\multirow{2}{*}{PhaseFormer} & 0.752 & 0.455 & 1.122 & 0.583 & 1.440 & 0.676 & 1.715 & 0.753 & 1.931 & 0.814 & 1.392 & 0.656 \\
 & $\pm$0.001 & $\pm$0.000 & $\pm$0.002 & $\pm$0.001 & $\pm$0.003 & $\pm$0.001 & $\pm$0.007 & $\pm$0.001 & $\pm$0.007 & $\pm$0.001 & $\pm$0.002 & $\pm$0.000 \\[1.6pt]
\multirow{2}{*}{TimeBridge} & 0.745 & 0.445 & 1.129 & 0.572 & 1.418 & 0.660 & 1.770 & 0.753 & 1.950 & 0.803 & 1.402 & 0.647 \\
 & $\pm$0.013 & $\pm$0.002 & $\pm$0.032 & $\pm$0.005 & $\pm$0.014 & $\pm$0.001 & $\pm$0.029 & $\pm$0.007 & $\pm$0.055 & $\pm$0.007 & $\pm$0.006 & $\pm$0.002 \\[1.6pt]
\multirow{2}{*}{TimeMixer++} & 0.726 & 0.443 & 1.089 & 0.565 & 1.394 & 0.659 & 1.647 & \underline{0.731} & 1.901 & 0.800 & 1.351 & 0.639 \\
 & $\pm$0.006 & $\pm$0.001 & $\pm$0.011 & $\pm$0.002 & $\pm$0.005 & $\pm$0.001 & $\pm$0.015 & $\pm$0.005 & $\pm$0.007 & $\pm$0.003 & $\pm$0.002 & $\pm$0.001 \\[1.6pt]
\midrule
\multirow{2}{*}{CANDOR (Ours)} & \textbf{0.712} & \textbf{0.440} & \textbf{1.040} & \textbf{0.559} & \textbf{1.302} & \textbf{0.644} & \textbf{1.542} & \textbf{0.715} & \textbf{1.763} & \textbf{0.780} & \textbf{1.272} & \textbf{0.628} \\
 & $\pm$0.001 & $\pm$0.000 & $\pm$0.004 & $\pm$0.003 & $\pm$0.024 & $\pm$0.005 & $\pm$0.011 & $\pm$0.001 & $\pm$0.027 & $\pm$0.014 & $\pm$0.012 & $\pm$0.004 \\[1.6pt]
\bottomrule
\end{tabular}
}
\end{table}

\subsection{Ablation Studies}
\label{abal}

\begin{figure*}[htbp]
\centering
\includegraphics[width=0.94\textwidth]{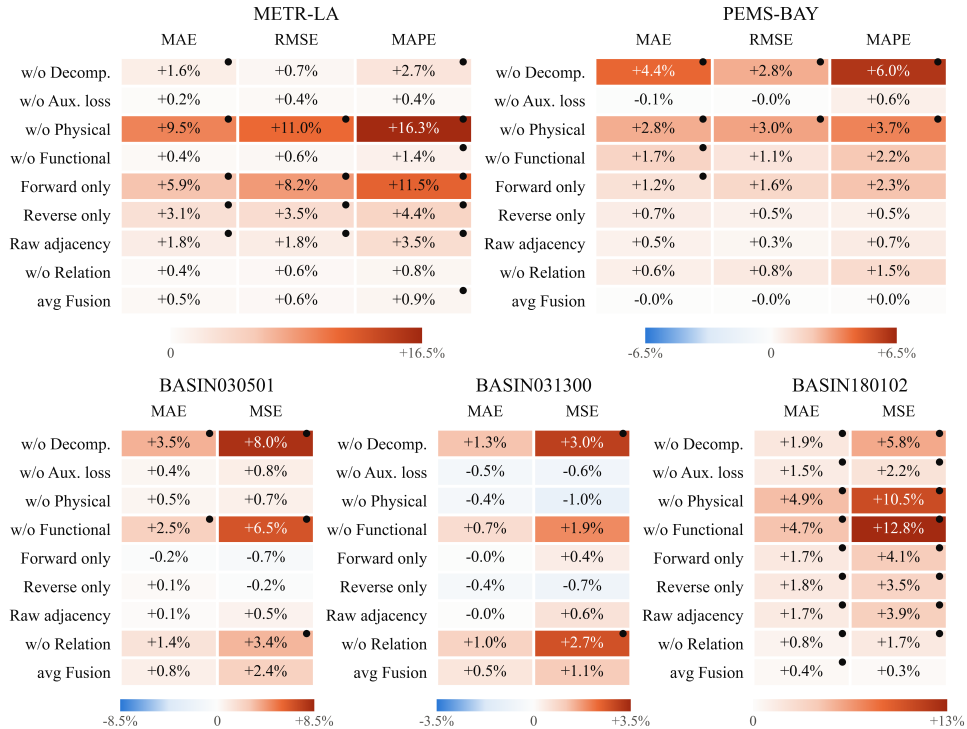}
\caption{{\color{black}Ablation variants across all five datasets. Values are
seed-paired relative changes (\%) from full CANDOR; higher is worse. Dots denote mean changes exceeding twice the full-model seed variation.}}
\label{fig:ablation_appendix}
\end{figure*}

{\color{black}
The complete study uses the same ten controlled variants and
seed-paired protocol as the main paper, extending the evaluation to all
five datasets. Figure~\ref{fig:ablation_appendix} reports relative error
changes from full CANDOR. The relational ablations reveal a systematic shift across systems.
Physical propagation is the dominant relational mechanism in
traffic: removing it increases MAE by
$9.5\%$ on METR-LA and $2.8\%$ on PEMS-BAY. In river networks,
functional relations become more influential. Their removal increases
MAE/MSE by $2.5\%/6.5\%$ on BASIN030501, while relation-specific
modeling contributes $2.7\%$ in MSE on BASIN031300. On BASIN180102,
removing the physical and functional branches increases MSE by
$10.5\%$ and $12.8\%$, respectively, showing that both mechanisms can
remain important within the same system. This changing branch
importance supports using complementary relation mechanisms instead of
a single graph shared across propagation regimes.

The remaining variants validate how these mechanisms are constructed
and connected. Both single-direction variants degrade performance where
physical propagation is informative, confirming distinct roles for
forward propagation and reverse support. Replacing the learned operator
with raw adjacency increases METR-LA MAPE by $3.5\%$ and BASIN180102
MSE by $3.9\%$, showing that topology alone cannot capture
temporal-offset and history-conditioned interactions. The ``D$^2$STGNN Physical'' replacement increases METR-LA
MAE by $1.0\%$ and yields comparable performance on PEMS-BAY.
At the 192-hour horizon, MSE increases by $1.1\%$, $0.3\%$, and
$6.6\%$ on BASIN030501, BASIN031300, and BASIN180102, respectively.
Removing
decomposition increases METR-LA MAE by $1.6\%$ and BASIN180102 MSE by $5.8\%$, with the largest effect on PEMS-BAY at $4.4\%/6.0\%$ in
MAE/MAPE; removing only its auxiliary supervision increases
BASIN180102 MAE/MSE by $1.5\%/2.2\%$. Relation-specific modeling and
adaptive fusion further contribute up to $3.4\%$ and $2.4\%$ in MSE.
Overall, the five-dataset matrix shows that CANDOR's gains do not arise
from a single dominant module. Decomposition, direction-aware
propagation, non-topological relations, and adaptive fusion contribute
under different structural regimes, allowing one architecture to
generalize across dense traffic networks and sparse river systems.

}


\subsection{Detailed Mechanism Analysis}
\label{app:mechanism_analysis}
This section examines whether CANDOR's gains follow from its intended
mechanisms beyond forecasting accuracy. We first ground the learned
Shock using rainfall unseen by CANDOR, verify that rainfall is
forecasting-relevant, and test whether it can supplement or replace the
Shock pathway. We then assess whether the decomposition is
information-preserving and role-specific, trace release and accumulation
through delayed physical propagation, and examine how
background-conditioned functional relations complement topology under
adaptive fusion. These analyses evaluate the physical
interpretability and end-to-end coherence of decomposition-conditioned relation learning.

\paragraph{External grounding of the learned Shock.}
We examine whether the learned decomposition acquires physical meaning
beyond its training objectives. Because rainfall can induce abrupt
changes in river conditions, we use observed precipitation as an
external probe of the Shock component. CANDOR never observes rainfall
during training, model selection, or standard forecasting. After
training, we compare station-aligned hourly precipitation from NASA GPM
IMERG V07 with the mean Shock-gate activity
$(g^{S}_{U,i,p}+g^{S}_{V,i,p})/2$, the same quantity regularized by
$\mathcal{L}_{\mathrm{shock}}$ in
Eq.~\ref{eq:app_component_regularization}. We compute Spearman correlations over lags from $-120$ to $+120$ hours
at two-hour intervals. Patch-level gate activities are averaged over
the patches covering each hour. Node-level correlations are averaged
within each seed and forecasting horizon, after which the maximum
correlation is selected for each horizon and averaged across horizons.
Statistical significance is assessed using 20,000 basin-level circular
shifts that preserve rainfall's temporal structure and repeat the
complete lag-selection procedure. Table~\ref{tab:shock_rainfall} shows significant positive associations
in all three basins. The selected correlation is also positive for all
45 seed-horizon checkpoints. Although the magnitudes vary across
systems, their consistent excess over the shift-based null provides
external physical grounding for the learned Shock role.

\begin{table}[htbp]
\centering
\caption{Association between the learned Shock-gate activity and
rainfall unobserved by CANDOR. The statistic is the mean horizon-wise
maximum Spearman correlation. The null distribution uses 20,000
circular shifts with the same lag-selection procedure.}
\label{tab:shock_rainfall}
\vspace{2pt}
\scriptsize
\setlength{\tabcolsep}{23pt}
\begin{tabular}{lrrrr}
\toprule
\textbf{Dataset} &
$\boldsymbol{\rho}$ &
\textbf{Null mean} &
\textbf{Excess} &
$\boldsymbol{p}$ \\
\midrule
BASIN030501 & 0.0347 & 0.0113  & 0.0234 & 0.0068 \\
BASIN031300 & 0.0445 & 0.0149  & 0.0296 & $<5\times10^{-5}$ \\
BASIN180102 & 0.1310 & -0.0115 & 0.1425 & 0.0181 \\
\bottomrule
\end{tabular}
\end{table}

\paragraph{Predictive relevance of rainfall.}
To assess whether rainfall carries predictive information, we evaluate
TimeXer~\citep{wang2024timexer} and XLinear~\citep{chen2026xlinear}, both of which support
exogenous inputs, with and without aligned rainfall. We report these
auxiliary comparisons separately because the rainfall-assisted models
receive an additional input unavailable to CANDOR and the standard
baselines. Table~\ref{tab:rainfall_aux} shows modest but consistent
improvements across all three basins, confirming the predictive
relevance of rainfall.

\begin{table*}[htbp]
\centering
\caption{TimeXer and XLinear with and without rainfall input. Results
are averaged over three seeds and five forecasting horizons.}
\label{tab:rainfall_aux}
\vspace{2pt}
\scriptsize
\setlength{\tabcolsep}{6pt}
\begin{tabular}{llcccccc}
\toprule
& & \multicolumn{2}{c}{\textbf{BASIN030501}}
& \multicolumn{2}{c}{\textbf{BASIN031300}}
& \multicolumn{2}{c}{\textbf{BASIN180102}} \\
\cmidrule(lr){3-4}
\cmidrule(lr){5-6}
\cmidrule(lr){7-8}
\textbf{Method} & \textbf{Rainfall}
& MAE & MSE
& MAE & MSE
& MAE & MSE \\
\midrule
TimeXer & No
& 0.4850 & 0.5883
& 0.4415 & 0.4280
& 0.6506 & 1.3794 \\
TimeXer & Yes
& \textbf{0.4835} & \textbf{0.5855}
& \textbf{0.4379} & \textbf{0.4231}
& \textbf{0.6493} & \textbf{1.3764} \\
\addlinespace
XLinear & No
& 0.4857 & 0.5926
& 0.4428 & 0.4333
& 0.6525 & 1.3902 \\
XLinear & Yes
& \textbf{0.4855} & \textbf{0.5923}
& \textbf{0.4422} & \textbf{0.4323}
& \textbf{0.6524} & \textbf{1.3899} \\
\bottomrule
\end{tabular}
\end{table*}

\paragraph{Intervening on the Shock pathway.}
We finally test whether rainfall adds to, or can replace, the learned
Shock representation. At the 192-hour horizon, \emph{RainAdd} injects
aligned rainfall into the native Shock stream, whereas \emph{RainOnly}
replaces that stream with rainfall. Both variants follow the same configuration and the same three seeds as full CANDOR. Table~\ref{tab:rainshock} shows that adding rainfall yields no distinguishable improvement, indicating that
its predictive contribution substantially overlaps with the learned
Shock representation. Rainfall alone provides no consistent advantage
and degrades MSE on BASIN180102 beyond the paired-variability criterion. These analyses show that CANDOR recovers rainfall-related
disturbances without observing rainfall, while the learned Shock
retains predictive information beyond precipitation alone.

\begin{table}[htbp]
\centering
\caption{Rainfall interventions at the 192-hour horizon. Values are
seed-paired relative changes (\%) from full CANDOR.}
\label{tab:rainshock}
\vspace{2pt}
\scriptsize
\setlength{\tabcolsep}{10pt}
\begin{tabular}{lrrrr}
\toprule
& \multicolumn{2}{c}{\textbf{RainAdd}}
& \multicolumn{2}{c}{\textbf{RainOnly}} \\
\cmidrule(lr){2-3}
\cmidrule(lr){4-5}
\textbf{Dataset} &
$\Delta$MAE & $\Delta$MSE &
$\Delta$MAE & $\Delta$MSE \\
\midrule
BASIN030501 & -0.33 & -1.05 & -0.78 & -1.04 \\
BASIN031300 & -1.15 & -1.86 & -0.67 & -0.65 \\
BASIN180102 & +0.50 & +0.49 & -0.16 & +1.77$^{\dagger}$ \\
\bottomrule
\end{tabular}

\vspace{2pt}
{\footnotesize $^{\dagger}$Change exceeds the paired-variability
criterion in Appendix~\ref{abal}.}
\end{table}

\paragraph{Reference-consistent decomposition.}
Figure~\ref{fig:mechanism_compact}(a) evaluates two complementary
properties of the learned decomposition: reconstruction sufficiency and
role differentiation. The current-reference reconstruction
\begin{equation}
\mathbf{Z}_{i,p}
\approx
\widetilde{\mathbf{B}}_{i,p}
+
\widetilde{\mathbf{A}}_{i,p}
+
\widetilde{\mathbf{S}}_{i,p}
\end{equation}
explains the observed state relative to its current background, making
$\widetilde{\mathbf{A}}_{i,p}$ responsible for gradual buildup above
that reference. The shifted-reference reconstruction
\begin{equation}
\mathbf{Z}_{i,p}
\approx
\widetilde{\mathbf{B}}_{i,\min(p+1,P)}
+
\widetilde{\mathbf{R}}_{i,p}
+
\widetilde{\mathbf{S}}_{i,p}
\end{equation}
instead measures the state relative to the subsequent background,
providing a complementary view of relaxation toward the next reference.
The shock component is shared by both reconstructions so that abrupt
deviations need not be absorbed into the smoother accumulation or
release components.

The agreement between the reconstructed and observed trajectories
shows that decomposition preserves information required to represent
the local dynamics. Meanwhile, the differentiated component regions in
the shared representation space indicate that reconstruction is not
achieved by four interchangeable copies of the same signal. These two
diagnostics provide complementary evidence: reconstruction assesses
whether the components remain collectively sufficient, whereas
representation separation assesses whether they acquire specialized
roles.

\paragraph{Delayed physical propagation and role consistency.}
For visualization, we combine the two directional propagation paths as
\begin{equation}
\mathbf{E}^{\Sigma}
=
\mathbf{E}^{\rightarrow}
+
\mathbf{E}^{\leftarrow},
\qquad
\mathbf{o}^{\Sigma}
=
\mathbf{o}^{\rightarrow}
+
\mathbf{o}^{\leftarrow},
\end{equation}
where $\mathbf{E}^{\Sigma}$ is the propagation residual and
$\mathbf{o}^{\Sigma}$ is the resulting incoming graph contribution.
Figure~\ref{fig:mechanism_compact}(b) examines their temporal ordering.
Under the two-patch propagation window used by CANDOR, the localized
states $\mathbf{E}^{\Sigma}(t)$ and
$\mathbf{E}^{\Sigma}(t+1)$ map to the later graph contribution
$\mathbf{o}^{\Sigma}(t+1)$. This ordering is consistent with delayed
transport: a deviation must first form a propagating state before its
effect appears as incoming information at another node.

Figure~\ref{fig:mechanism_compact}(c) examines whether this transition
preserves the intended component semantics. Release
$\mathbf{R}_{i,p}$ aligns with the contemporaneous propagation residual
$\mathbf{E}^{\Sigma}_{i,p}$, while subsequent accumulation
$\mathbf{A}_{i,p+1}$ aligns with the incoming contribution
$\mathbf{o}^{\Sigma}_{i,p}$. Thus, release-associated deviations provide
states that can be transmitted, whereas the received graph contribution
supports later buildup at the receiving node. These alignments are
learned representation-level correspondences rather than assertions of
exact mass conservation. Likewise, the reverse-support path provides
counter-directional predictive context and should not be interpreted as
backward physical transport.

\paragraph{Background-conditioned relations beyond topology.}
Figure~\ref{fig:system_analysis}(a) compares the directed physical
support $\mathcal{G}_0$ with the learned functional relations for a
representative water-quality site. Several nodes outside the physical
neighborhood receive substantial functional weights, showing that
predictive dependence is not restricted to direct river connectivity.
Such links can arise when distant sites share persistent temporal
behavior even though no direct transport edge connects them.

Importantly, these relations are inferred from background dynamics
$\mathbf{B}$ rather than from unrestricted mixtures of all components.
Persistent background behavior provides a more stable basis for
relation discovery, while transient accumulation, release, and shock
events are prevented from directly redefining the functional graph.
A functional edge should therefore be interpreted as a predictive
relationship, not as evidence of an unobserved physical connection or
causal transport pathway. Its role is to complement, rather than
replace, the physically supported graph.

\paragraph{Context-adaptive use of relation mechanisms.}
Figure~\ref{fig:system_analysis}(b) reports both the mean allocation and
the $10$th-$90$th percentile range of the fusion weights across nodes,
samples, and forecast positions. The nonzero ranges show that fusion is
not merely dataset-specific: the relative use of functional, forward,
and reverse-support forecasts also changes within each dataset and
across forecasting contexts. The allocation patterns are consistent with the different propagation
regimes. METR-LA assigns greater average weight to physical support,
where directed interactions provide useful information about congestion
propagation. PEMS-BAY and the water-quality datasets assign relatively
more weight to functional forecasts, indicating that persistent
dependencies beyond immediate topology become more useful when physical
coupling is weaker, sparser, or subject to heterogeneous delays. No
single branch dominates uniformly. Together with the nonlocal relations
in panel (a), this result supports the central design of CANDOR:
physical propagation and functional dependence capture distinct sources
of predictive information, and their relevance must be adapted to the
observed system and forecasting context.

\end{document}